\documentclass[letterpaper]{article} 
\usepackage{aaai2026}  
\usepackage{times}  
\usepackage{helvet}  
\usepackage{courier}  
\usepackage[hyphens]{url}  
\usepackage{graphicx} 
\usepackage{natbib}  
\usepackage{caption} 
\usepackage{algorithm}
\usepackage{algorithmic}

\usepackage{newfloat}
\usepackage{listings}
\DeclareCaptionStyle{ruled}{labelfont=normalfont,labelsep=colon,strut=off} 
\floatstyle{ruled}
\newfloat{listing}{tb}{lst}{}
\floatname{listing}{Listing}
\usepackage{algorithm}
\usepackage{algorithmic}
\usepackage{amsmath}
\usepackage{amsfonts}
\usepackage{subcaption}
\usepackage{multirow}
\usepackage{booktabs}
\usepackage{makecell}
\usepackage{nicefrac}
\usepackage{graphicx}
\usepackage{pifont} 
\usepackage{bm}
\newcommand{\cmark}{\ding{51}} 
\newcommand{\xmark}{\ding{55}} 
\usepackage{breqn}
\newif\ifincludeappendix
\includeappendixtrue

\title{VEDA: Generation of 3D Molecules via Variance-Exploding Diffusion \\ with Annealing}
\author{
    Peining Zhang\textsuperscript{\rm 1}, Jinbo Bi\textsuperscript{\rm 1}, Minghu Song\textsuperscript{\rm 2}
}
\affiliations{
    \textsuperscript{\rm 1}University of Connecticut, Storrs, Connecticut 06269, USA\\
    \textsuperscript{\rm 1}Institute of Health and Medicine, Hefei Comprehensive National Science Center, Hefei 230601, China\\
    	
    peining.zhang@uconn.edu,
    jinbo.bi@uconn.edu,
    minghu.song@ihm.ac.cn
}

\usepackage{bibentry}

\begin{document}

\maketitle

\begin{abstract}
Diffusion models show promise for 3D molecular generation, but face a fundamental trade-off between sampling efficiency and conformational accuracy. While flow-based models are fast, they often produce geometrically inaccurate structures, as they have difficulty capturing the multimodal distributions of molecular conformations. In contrast, denoising diffusion models are more accurate but suffer from slow sampling, a limitation attributed to sub-optimal integration between diffusion dynamics and SE(3)‑equivariant architectures. 
To address this, we propose \textbf{VEDA}, a unified SE(3)-equivariant framework that combines variance-exploding diffusion with annealing to efficiently generate conformationally accurate 3D molecular structures. Specifically, our key technical contributions include: (1) a VE schedule that enables noise injection functionally analogous to simulated annealing, improving 3D accuracy and reducing relaxation energy; (2) a novel preconditioning scheme that reconciles the coordinate-predicting nature of SE(3)-equivariant networks with a residual-based diffusion objective, and (3) a new arcsin-based scheduler that concentrates sampling in critical intervals of the logarithmic signal-to-noise ratio. 
On the QM9 and GEOM-DRUGS datasets, VEDA matches the sampling efficiency of flow-based models, achieving state-of-the-art valency stability and validity with only 100 sampling steps. More importantly, VEDA's generated structures are remarkably stable, as measured by their relaxation energy ($\Delta E_{\text{relax}}$) during GFN2-xTB optimization. The median energy change is only 1.72~kcal/mol, significantly lower than the 32.3~kcal/mol from its architectural baseline, SemlaFlow. Our framework demonstrates that principled integration of VE diffusion with SE(3)-equivariant architectures can achieve both high chemical accuracy and computational efficiency.
\end{abstract}

\begin{links}
    \link{Code}{https://github.com/peiningzhang/VEDA}
    \link{Extended version}{https://arxiv.org/abs/2511.09568}
\end{links}
\section{Introduction}
\label{sec:introduction}

Deep generative models are revolutionizing computational drug discovery. Breakthroughs like AlphaFold3~\cite{abramson2024accurate} and RFdiffusion~\cite{watson2023novo} have shown the power of AI in protein structure prediction and design. Among them, diffusion models \cite{ho2020denoising, songscore} and their variants~\cite{lipman2023flow}, originally developed for computer vision, have emerged as a dominant tool for generating novel small molecules, a core task in drug discovery.  Their ability to model continuous generative processes (i.e., transforming random noise into complex, structured data through a series of continuous steps) well aligns with the continuous coordinates of 3D molecular conformation.  By representing molecules as 3D point clouds in Euclidean space and using equivariant architectures (e.g., equivariant graph neural networks (EGNNs)~\cite{satorras2021n_EGNN} and E(3)-equivariant Transformers~\cite{fuchs2020se_Etransformer},) these models can generate geometrically realistic and chemically valid conformations~\cite{hoogeboom2022equivariant}.

In recent years, numerous diffusion models have been developed for 3D molecular generation or 3D molecular conformation generation. However, existing approaches have largely focused on refining their network architecture or by injecting domain-specific knowledge to guide the generation process—for instance, by developing novel equivariant transformers~\cite{liaoequiformer}, shifting from direct 3D atom coordinates to internal coordinates like torsion angles~\cite{jing2022torsional}, or explicitly modeling chemical bond interactions~\cite{huang2023mdm}—while overlooking the principled design of the diffusion process itself~\cite{zhang2025unraveling}. A critical yet unaddressed question is how to redesign the learning objective to best match a unique characteristic of SE(3)-equivariant neural networks: their strong tendency to learn identity-like mappings, a bias stemming from their message-passing mechanism and strict geometric constraints. In diffusion models, this is often achieved through a technique called preconditioning, which provides a vital analytical framework to make the training objective more tractable~\cite{karras2022elucidating}. By aligning the objective with the model's architectural inductive biases, this principled approach can significantly improve performance and training stability. Despite its potential, this promising direction remains largely overlooked in current molecular generative models.

To address this gap, we propose VEDA, a principled framework for 3D molecular diffusion where:
\begin{itemize}
\item To the best of our knowledge, VEDA is the first framework to apply the Variance-Exploding (VE) diffusion paradigm to the hybrid discrete-continuous domain of 3D molecules, unifying atomic types (discrete) and coordinates (continuous) within a single diffusion process that is functionally analogous to simulated annealing.

\item We propose a theoretically grounded preconditioning scheme to correct the inductive bias of coordinate-predicting SE(3) equivariant networks. As a complementary innovation in diffusion dynamics, we introduce a noise schedule based on the arcsine function, which achieves a better balance between early-stage exploration and late-stage refinement in molecular generation.

\item VEDA substantially reduces relaxation energy of generated molecules, by 90\% compared to those generated by SemlaFlowa~\cite{irwin2025semlaflow}, while achieving state-of-the-art performance on QM9 and GEOM-DRUGS, and matching the efficiency of strong flow-based models with significantly fewer sampling steps.
\end{itemize}

\begin{figure*}[htb]
    \centering
    \includegraphics[width=1\linewidth, trim=3.5cm 8cm 3.5cm 1cm, clip]{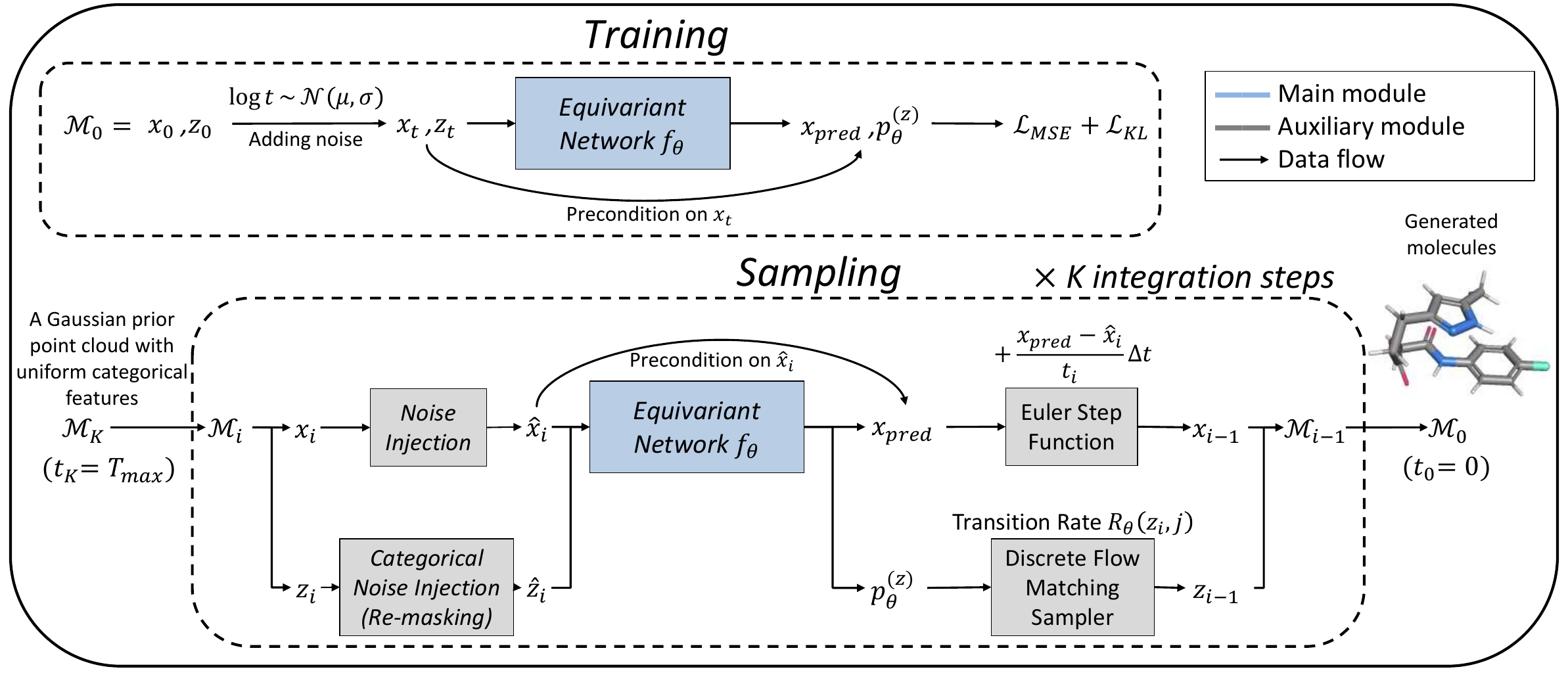}
    \caption{
        An overview of the VEDA framework, detailing its training and sampling processes. 
        During \textbf{Training} (top), a clean molecule $\mathcal{M}_0$ is perturbed via Gaussian noise for coordinates ($\mathbf{x}_t$) and categorical corruption for features ($\mathbf{z}_t$), defined by Eq.~\ref{eq:coodinate_definition} and Eq.~\ref{eq:type_definition}.
        The equivariant network $f_\theta$ is then trained to predict the original molecule by minimizing the combined Mean Squared Error (MSE) and Cross-Entropy (CE) loss in Eq.~\ref{eq:Loss_veda}.
        During \textbf{Sampling} (bottom), the process starts from a pure noise distribution (a Gaussian point cloud with uniform categorical features) and iteratively refines the sample over $K$ steps. 
        Each integration step $i$ involves: 
        (1) a noise injection from $(\mathbf{x}_i,\mathbf{z}_i)$ to $(\mathbf{\hat{x}}_i, \mathbf{\hat{z}}_i)$; 
        (2)  the network $f_\theta$ is applied to both coordinates and features, with preconditioning affecting only the coordinate predictions; it outputs $x_{\text{pred}}$ and the category probabilities $p^{(z)}_\theta$ (Eq.~\ref{eq:updated_precondition}); 
        and (3) an update combining a continuous Euler step and a Discrete Flow Matching sampler to obtain $\mathcal{M}_{i-1}$, formally given in Eq.~\ref{eq:step_function_coordinated} and Eq.~\ref{eq:transition_rate}.
        In the diagram, blue boxes represent the main network module, gray boxes are auxiliary operations, and arrows indicate the data flow.
    }
    \label{fig:model-diagram}
\end{figure*}

\section{Related works}
\label{sec:related_works}
\paragraph{Diffusion-Based Generative Models}
Denoising Diffusion Probabilistic Models (DDPMs)~\cite{ho2020denoising} have become a leading class of generative models which learn to reverse a forward process that incrementally corrupts data with Gaussian noise. 
DDPMs are extended to continuous-time settings using score-based stochastic differential equations (SDEs)~\cite{songscore}, incorporating either variance-preserving or variance-exploding dynamics.
Recent advances, particularly those utilizing Variance-Exploding (VE) SDEs with preconditioning techniques~\cite{karras2022elucidating}, have substantially enhanced generation quality and sampling efficiency.
An alternative approach, Flow Matching~\cite{lipman2023flow}, directly learns the velocity field of  the data distribution's probability flow, offering a different yet highly competitive approach to generative modeling.
Consistent with recent literature~\cite{gao2024diffusion}, we consider flow matching a specialized formulation within the broader diffusion framework. 

\paragraph{Discrete Diffusion Models}
The principles of diffusion have also been extended to discrete data generation, drawing inspiration from masked language modeling~\cite{devlin2019bert}.
These non-autoregressive methods are a natural fit for the unordered structure of molecules, enabling de novo generation without imposing artificial sequential or ordering assumptions.
D3PM~\cite{austin2021structured} established the theoretical foundations for discrete diffusion using continuous-time Markov chains.
Subsequent developments of variants of Discrete Flow Matching~\cite{campbell2024generative,gat2024discrete} have further advanced the field with improved sampling strategies and scalability.

\paragraph{\textit{De novo} 3D Molecular Generation}
Modern 3D molecular generation heavily relies on E(3)-equivariant neural networks, from pioneering architectures like EGNN~\cite{satorras2021n_EGNN} to subsequent transformer-based models~\cite{zhang2025d3mes}.
From a probabilistic modeling perspective, the field is dominated by a fundamental trade-off.
Denoising diffusion-based methods like EDM~\cite{hoogeboom2022equivariant} and GeoLDM~\cite{xu2023geometric} achieve high conformational accuracy but suffer from slow sampling.
Conversely, flow-based models such as EquiFM~\cite{song2023equivariantFM}, GeoBFN~\cite{song2023unified}, and SemlaFlow~\cite{irwin2025semlaflow} are efficient but often struggle with geometric precision.
Balancing high fidelity with computational efficiency remains a pressing challenge, with recent benchmarks underscoring the chemical inaccuracy of flow-based models~\cite{nikitin2025geom}.

\section{Methodology}
\label{sec:methodology}
\subsection{Preliminaries}
We represent molecules as point clouds of chemical elements in 3D space, denoted by $\mathcal{M} = (\mathbf{x}, \mathbf{z})$, where $\mathbf{x} = (\mathbf{x}_1, \dots, \mathbf{x}_N) \in \mathbb{R}^{N \times 3}$ are the atomic coordinates of $N$ atoms, and $\mathbf{z}$ denotes non-geometric molecular attributes.  
To keep notation unified, we decompose $\mathbf{z}=(\mathbf{h},\mathbf{b})$, where $\mathbf{h}$ includes atom-level properties (e.g., atom types, charges) and $\mathbf{b}$ denotes pairwise bond information.

In most graph-based models, a molecule is further represented by a graph $G = (V, E)$, where nodes $V$ represent atoms and edges $E$ represent either inferred or explicitly defined interactions between atoms.
These interactions can be constructed based on distance thresholds or chemical bond annotations. 
Some models treat bond information as part of the output (explicit bond modeling), while others infer it post hoc based on generated atomic coordinates (implicit bond inference).
This distinction influences the architecture and training strategy of generative models, as detailed later.

\subsection{Model Architecture}
To show our VEDA framework is both general and scalable, we apply it to two architectures at different levels of complexity. This dual implementation proves VEDA's effectiveness across different modeling paradigms, covering both implicit and explicit bond generation.
\noindent
\paragraph{VEDA-E: Implicit Bond Modeling} 
The VEDA-E variant is built upon the EGNN architecture~\cite{satorras2021n_EGNN} from EDM~\cite{hoogeboom2022equivariant}. In this setup, the model generates only node-centric features $\mathbf{h} = (\mathbf{h}_1, \dots, \mathbf{h}_N) \in \mathbb{R}^{N \times d_h}$, which include atom types and charges. Diffusion is performed by adding Gaussian noise to all continuous features. The chemical bond structure $\mathbf{b}$ is not explicitly modeled during generation; instead, it is inferred post hoc based on interatomic distances and chemical valence rules, a standard practice in EDM-based molecule generation.
\noindent
\paragraph{VEDA-S: Explicit Bond Modeling}
In contrast, VEDA-S adopts the Semla architecture from SemlaFlow~\cite{irwin2025semlaflow} to explicitly model bonds. This variant directly generates a complete molecular graph representation $\mathbf{z} = (\mathbf{h}, \mathbf{b})$, where $\mathbf{h}$ represents node features and $\mathbf{b} \in \mathbb{R}^{N \times N \times d_b}$ denotes the bond types between atoms. This approach is compatible with discrete data types by using a mask-based diffusion process and a classification objective. A diagram illustrating the main concept of VEDA-S is provided in Figure~\ref{fig:model-diagram}.

\subsection{Diffusion Dynamics}
Our generative model is based on Score Diffusion~\cite{songscore}. It includes a forward diffusion process gradually perturbing the clean data into noise, and a learned denoising process that recover the noise into data. The denoising process is governed by the denoiser $D_\theta$, which takes as input the noisy sample $\mathbf{x}_t$ and the noise level $t$. We define $\mathbf{x}_0$ as the clean data point (e.g., molecular coordinates) and $\mathbf{x}_{\infty}$ as the fully noised sample.
\noindent
\paragraph{Forward Process}
We corrupt the continuous coordinates $\mathbf{x}$ by adding Gaussian noise:
\begin{align}
\label{eq:coodinate_definition}
\mathbf{x}_t& = \mathbf{x}_0 + t\,\epsilon,\quad \epsilon{\sim}\mathcal{N}(0,I),
\end{align}
The noise level $t$ is sampled from a log-normal distribution: $t \sim \operatorname{LogNormal}\left(\ln\sqrt{T_{\min} T_{\max}}, [\frac{1}{8}\ln(T_{\max} / T_{\min})]^2\right)$, so $t\in [T_{\min} , T_{\max}])$ in most case. 

\noindent For discrete features $\mathbf{z}$, we consider two variants.

\noindent In \textbf{VEDA-E}, Gaussian noise is applied:
{\footnotesize
\begin{equation}
\mathbf{z}_t = \mathbf{z}_0 + t,\eta, \quad \eta \sim \mathcal{N}(0, I).
\end{equation}}
\noindent In \textbf{VEDA-S}, we instead use time-dependent categorical corruption:
{\footnotesize
\begin{equation}
\label{eq:type_definition}
\mathbf{z}_t \sim \mathrm{Cat}!\left[(1 - m(t)),\delta(\mathbf{z}_t = \mathbf{z}_0) + \frac{m(t)}{S}\right],
\end{equation}}
\noindent where \(S\) is the number of categories and
\begin{equation}
m(t)=\frac{t}{t+\sigma_d}
\end{equation}
is the categorical corruption rate, with \(\sigma_d\) denoting the data scale used in the continuous diffusion process. This schedule monotonically increases the contribution of uniform categorical noise as \(t\) increases, thereby coupling the discrete corruption level to the continuous VE noise level. At large \(t\), the categorical distribution approaches the uniform distribution.

Our approach uses this principled VE scheduler during training. This variance-exploding (VE) formulation is particularly suitable for modeling continuous data such as 3D molecular structures, as it enables flexible noise schedules and accurate noise level control in log-space, injecting a massive amount of noise to reach lower energy. We believe infinite separation at $t\to\infty$ between atoms is an ideal noisy state for molecule generation, superior to a standard Gaussian prior, as it corresponds to no chemical interactions between atoms. 
\noindent
\paragraph{Preconditioning the Denoiser}
In 3D molecular generation with diffusion models, directly predicting coordinates ($x$-prediction) is the prevailing approach, as equivariant network architectures like GNNs are not well-suited for predicting noise or velocity~\cite{irwin2025semlaflow}. We adopt the preconditioning framework from \citet{karras2022elucidating}, where the denoiser $D_\theta$ is defined as a combination of the input $\mathbf{x}_t$ and a neural network output $F_\theta$:
\begin{align}
    D_\theta(\mathbf{x}_t; t) &= c_{\text{skip}} \mathbf{x}_t +
    c_{\text{out}} \left[ F_\theta(c_{\text{in}} \mathbf{x}_t; c_{\text{noise}}) \right]\\
    F_{\text{target}} &= \frac{1}{c_{\text{out}}}(\mathbf{x}_0 - c_{\text{skip}}(\mathbf{x}_0 + t\,\epsilon))
\end{align}

Optimizing the preconditioning framework with a coordinate-space MSE loss requires the network ($F_\theta$) to output residuals that are uncorrelated with the input. However, our backbone (GNN or Graph Transformer) outputs coordinates via a residual connection design~\cite{he2016deep}, which causes $F_\theta$ to produce outputs highly correlated with $\mathbf{x}_t$, conflicting with the residual learning objective.

To address this mismatch, we subtract a scaled identity component from the network output. Instead of forcing the network to suppress its inherent bias through the loss function, we explicitly subtract  the undesired identity component from its output. Our modified denoiser is:
\begin{equation}
\label{eq:updated_precondition}
    D_\theta(\mathbf{x}_t; t) = c_{\text{skip}} \mathbf{x}_t + c_{\text{out}} \left( F_\theta(c_{\text{in}} \mathbf{x}_t; c_{\text{noise}}) - \alpha_t\,c_{\text{in}}\,\mathbf{x}_t \right)
\end{equation}
The coefficient $\alpha_t$ is our main contribution in this section. It serves as the optimal linear predictor of the ground-truth noise $\frac{\sigma_d \boldsymbol{\epsilon}}{\sqrt{\sigma_d^2 + t^2}}$ from $F_{\text{target}}$, where $\sigma_d$ is the standard deviation of the data distribution~\cite{karras2022elucidating}. This corresponds to the Linear Minimum Mean Squared Error (LMMSE) solution, which yields the following closed-form coefficient:
{\small
\begin{align}
\alpha_t &= \arg \min_\alpha\ \mathbb{E}[\| \frac{\sigma_d \boldsymbol{\epsilon}}{\sqrt{\sigma_d^2 + t^2}} - \alpha_t\,c_{\text{in}}\,\mathbf{x}_t \|^2] \\
&= \arg \min_\alpha\ \mathbb{E}[\| \sigma_d \boldsymbol{\epsilon} - \alpha_t \mathbf{x}_t \|^2] \\
&=\frac{\text{Cov}(\sigma_d \boldsymbol{\epsilon}, \mathbf{x}_t)}{\text{Var}(\mathbf{x}_t)} = \frac{\sigma_d\,t}{\sigma_d^2 + t^2}    
\end{align}}Using this optimal $\alpha_t$ reduces the residual correlation with the input and better aligns the training objective with the network’s inductive bias, i.e., its preference for modeling absolute coordinates.\ifincludeappendix
(See Appendix~\ref{appendix:alpha_opt_derivation} for derivation.)
\else
\fi 
We use the standard definitions of $c_{\text{skip}}$, $c_{\text{out}}$, $c_{\text{in}}$, $c_{\text{noise}}$, and $\lambda_\sigma$ from \citet{karras2022elucidating}.


\subsection{Training Objectives}
For VEDA-E,  we apply a mean squared error (MSE) loss to both continuous coordinates and categorical features:
\begin{align}
\mathcal{L}_{\text{MSE}} &= \mathbb{E}_{\mathbf{x}_0, \mathbf{z}_0, \epsilon, t} \left[ \left\| D_\theta(\mathbf{x}_t, \mathbf{z}_t; t) - (\mathbf{x}_0, \mathbf{z}_0) \right\|^2 \right],
\end{align}
where $(\mathbf{x}_0, \mathbf{z}_0)$ represents the concatenated true coordinates and categorical features. 
Following the implementation of EDM~\cite{hoogeboom2022equivariant}, VEDA-E treats categorical features $\mathbf{z}_0$ as continuous during denoising and applies MSE loss, despite their inherently discrete nature.. 

For VEDA-S, the model predicts the original discrete features $\mathbf{z}_0$ via a categorical distribution $\hat{p}_\theta(\mathbf{z}_0 \mid \tilde{\mathbf{z}}_t, t)$. This is optimized using a KL divergence loss:
\begin{equation}
\mathcal{L}_{\mathrm{KL}}
= \mathbb{E}_{\mathbf{z}_0,\mathbf{\tilde z_t},t}\Bigl[\mathrm{KL}\bigl(\delta(\mathbf{z}_0)\,\|\,\hat p_\theta(\mathbf{z}_0\mid \mathbf{\tilde z_t},t)\bigr)\Bigr].
\end{equation} 
The overall loss combines continuous and discrete terms:
\begin{equation}
\label{eq:Loss_veda}
\mathcal{L}_{\text{VEDA-S}} = \lambda_{\mathrm{cont}}\;\mathcal{L}_{\mathrm{MSE}} \;+\; \lambda_{\mathrm{disc}}\;\mathcal{L}_{\mathrm{KL}}
\end{equation}
where $\lambda_{\mathrm{cont}}$ and $\lambda_{\mathrm{disc}}$ balance the two objectives.

\subsection{Sampling}
We adopt the standard denoising framework where the generative process iteratively refines noisy inputs into data samples, following a predefined SDE  formulation~\cite{karras2022elucidating}. Conceptually, this process is analogous to \emph{simulated annealing}. The noise level \(t\) acts as a temperature parameter that is gradually lowered, enabling broad exploration of the energy landscape at high noise and convergence to low-energy, stable conformations as noise decreases. Our model operates under the $x$-prediction parameterization to directly estimate the clean sample from its noisy counterpart. During sampling, VEDA-E applies continuous noise injection to all features, while VEDA-S uses continuous noise injection for coordinates and discrete token refinement for categorical features. We now describe three interlocking sampling components: (1) continuous denoising with amplified noise injection, (2) discrete masked-token refinement, and (3) the proposed arcsin noise scheduler.

\paragraph{Stochastic Annealing for Continuous Coordinates}
Our sampling process employs an amplified noise injection strategy. This approach is governed by a hyperparameter, $\gamma>0$, which controls the degree of noise amplification. This process is equivalent in expectation to Gaussian smoothing on the molecular potential energy surface with a Gaussian kernel bandwidth of $\sqrt{\gamma^2 + 2\gamma} \cdot t_i$ \ifincludeappendix
(See Appendix~\ref{appendix:smoothing_derivation} for deviation)
\else
\fi. This pronounced smoothing suppresses local minima and surface roughness, making it easier for the sample trajectory to find the global low-energy basin and thereby improving the final energy metric~\cite{miao2015gaussian}. 

Sampling proceeds in two substeps for each iteration $i$:
\begin{enumerate}
\item \textbf{Perturbation}: We first set the value of $\hat{t}_i=(1+\gamma)t_i$ inject amplified noise into the current sample $\mathbf{x}_i$ to reach an intermediate state $\hat{\mathbf{x}}_i = \mathbf{x}_i + \sqrt{\hat{t}_i^2 - t_i^2} \cdot \boldsymbol{\epsilon}$, where $\boldsymbol{\epsilon} {\sim} \mathcal{N}(0, I)$. The noise added is substantially greater than in standard DDPM~\cite{ho2020denoising} or Flow-Matching~\cite{lipman2023flow} schedules.
\item \textbf{Denoising \& Update}: The denoiser \(D_\theta(\hat{\mathbf{x}}_i; \hat{t}_i)\) predicts the clean structure, from which we extrapolate the next sample $\mathbf{x}_{i+1}$ via a Euler step:
{\small
\begin{equation}
\label{eq:step_function_coordinated}
\mathbf{x}_{i+1} = \hat{\mathbf{x}}_i + \frac{t_{i+1} - \hat{t}_i}{\hat{t}_i}\bigl(\hat{\mathbf{x}}_i - D_\theta(\hat{\mathbf{x}}_i; \hat{t}_i)\bigr).
\end{equation}}
\end{enumerate}
In VEDA-E, this two-step procedure is applied uniformly to both coordinates and categorical features. In VEDA-S, only the coordinates undergo amplified noise injection, while categorical features are handled separately via discrete token refinement.

\paragraph{Masked Token Refinement for Discrete Variables}
For discrete features in VEDA-S, we use masked token refinement process. This process is governed by the time-dependent mask rate function $m(t)$, which is designed to align the discrete corruption with the continuous noise schedule. Specifically, when injecting noise from noise level $t_1$ to $t_2$, each token is randomly re-masked with probability $\frac{m(t_2)-m(t_1)}{1-m(t_1)}$. Additionally, we found that a simple uniform random re-masking strategy consistently outperformed more complex confidence-based strategies proposed in prior work~\cite{nie2025large}. We included our comparison with the strategies of low prediction confidence~\cite{chang2022maskgit} or small probability margins~\cite{kimtrain} \ifincludeappendix
in Appendix~\ref{appendix:low_conf_remask}.
\else
.
\fi

For discrete sampling, our approach is a step-wise masked sampling algorithm based on Continuous-Time Markov Chains (CTMC) within the Discrete Flow Matching framework. We implement and compare two complementary sampling strategies based on this foundation. 

The full transition rate combines base interpolation dynamics with detailed balance corrections:
\begin{equation}
\label{eq:transition_rate}
    R_{\theta}(z_t,j) = \omega(t) \cdot p_\theta(z_0 = j|z_t) + \eta_t \cdot p_\theta(z_0=z_t|z_t)
\end{equation}
where $\omega(t)=\frac{\eta_t S(1- m(t)) + \eta_t m(t) +m'(t)}{m(t)}$ is the time-dependent scaling factor, and $\eta_t$ is based on the categorical noise hyperparameter $\eta$. To determine the optimal configuration, we performed a comprehensive grid search. The results confirmed that the variable setting from \citet{campbell2024generative}, where $\eta_t=\eta m'(t)$, achieves the best and most robust performance, outperforming both the fixed setting ($\eta_t=\eta$) and the simplified DFM formulation from \citet{gat2024discrete}. The transition rates are thus derived to optimally combine the model's confidence in clean data predictions with the stability of maintaining current token assignments\ifincludeappendix
(see Appendices~\ref{appendix:mask_transition_rate}, \ref{appendix:gat_dfm_mask_transition_rate} for derivations and~\ref{appendix:ablation_discrete_sampling} for grid search details).
\else
\fi

\begin{figure}[h]
    \centering
    \includegraphics[width=1\linewidth, trim=0cm 0cm 0cm 0cm, clip]{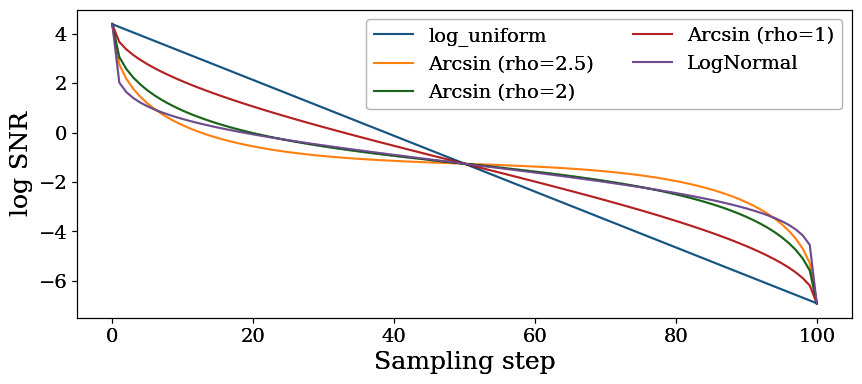}
    \caption{The arcsin sampling scheduler is proposed to focus on the middle part of sampling where log(SNR) is close to 0}
    \label{fig:scheduler_comparison}
\end{figure}

\paragraph{Proposed Noise Scheduler}
We propose an \textbf{arcsin-based noise scheduler} that further improves the sampling process. Through empirical analysis, we observe that sampling steps corresponding to near-zero log signal-to-noise ratio (log-SNR), i.e. $\log(\text{signal variance/noise variance})$ are especially critical for final molecular structure formation. It aligns well with the distribution of $t \sim \operatorname{LogNormal}\left(\ln\sqrt{T_{\min} T_{\max}}, [\frac{1}{8}\ln(T_{\max} / T_{\min})]^2\right)$ used during training.
To leverage this observation, let $u=i/N$ is the normalized step index for a total of $N$ steps. We design an arcsin-shaped scheduler parameterized by a tunable scalar $\rho$. Our scheduler is defined as:
{\footnotesize
\begin{align}
w(u) &= (1-\rho)\,u \;+\;\rho\,\tfrac{2}{\pi}\arcsin(\sqrt{u})\\ 
t^{(i)} &= T_{\min}\Bigl(\tfrac{T_{\max}}{T_{\min}}\Bigr)^{w(u)}
\end{align}
}
\noindent Where $\rho \in [0, \tfrac{\pi}{\pi-2}]$ modulates the concentration around log-SNR $\approx 0$. When $\rho = 0$, the scheduler reduces to a log-uniform schedule commonly used in prior work; as $\rho$ increases, more steps cluster in the mid-range. This targeted allocation of sampling steps improves structural fidelity and overall sample quality, especially in chemically sensitive configurations. 
As shown in Figure~\ref{fig:scheduler_comparison}, the arcsin scheduler (\(\rho{=}2\)) closely matches the log-normal training distribution, further supporting its design rationale.

\section{Experiments}

\begin{table}[htb]
\small
\centering
\setlength{\tabcolsep}{1mm}
\begin{tabular}{l|c|cccc}
\hline
Methods          & NFE ↓ & \makecell{Atom\\Sta(\%)↑} & \makecell{Mol\\Sta(\%)↑} & \makecell{Valid\\(\%)↑} & \makecell{Valid \&\\Unique(\%)↑}\\
\hline
EDM & 1000 
&98.7&82&91.9&90.9 \\
GeoLDM & 1000 
&98.9&89.4&93.8&92.6 \\
UniGEM & 1000 
&99.0& 89.8& 95& 93.2\\
EquiFM & 210 
&98.9&88.3&94.7&93.5  \\
GeoBFN & 2000
& 99.3& 93.3& 96.9& 92.4  \\
GOAT & 90 
& 99.2 & - & 92.9& 92.0\\
VEDA-E & \textbf{50}& \textbf{99.4}&\textbf{93.7}&\textbf{98.1}&\textbf{97.9}\\
VEDA-E & \textbf{30}& 99.2&91.5&97.0&96.8\\
\hline
MiDi & 500
& 97.5& 97.9& 95.5& - \\
FlowMol2 & 100
& 99.9& 99.6& 99.5& - \\
EQGAT-diff & 500
& 99.9& 98.7& 99.0& 98.8\\
SemlaFlow & 100
& 99.9& 99.7& 99.4&98.7\\
VEDA-S & 100
& \textbf{100.0}& \textbf{99.9}& \textbf{99.6}& \textbf{98.9} \\
VEDA-S & 50
& \textbf{100.0}& 99.7& 99.4& 98.4 \\
VEDA-S & 30
& 99.9& 99.5& 99.5& 98.9 \\
\hline
\end{tabular}
\caption{
\textbf{Results on QM9.}
\emph{Top:} bond‑implicit methods (bonds inferred from coordinates);  
\emph{Bottom:} bond‑explicit methods (bonds generated directly).  
NFE = number of function evaluations (sampling steps).  
Best in each group in \textbf{bold}.  }
\label{tab:uncond_qm9}
\end{table}

\subsection{Experimental Setup}
\paragraph{Model Variants}
To evaluate our VE-diffusion paradigm, we implement two variants based on distinct backbones. \textbf{VEDA-E} is built upon the EDM framework~\cite{hoogeboom2022equivariant}, using an EGNN backbone and performing implicit bond modeling. In contrast, \textbf{VEDA-S} derives from the SemlaFlow framework~\cite{irwin2025semlaflow}, adopting a Semla backbone for explicit bond modeling.


\begin{table*}[ht]
\centering
\small
\resizebox{\textwidth}{!}{
    \begin{tabular}{l|cc|cc|ccc|ccll}
        \hline
        Model & NFE & \makecell{Time\\(s)} & MS$\uparrow$ & \makecell{V\&C}$\uparrow$ & \makecell{Bond\\Length\\($\times10^{-2}$)}$\downarrow$ & \makecell{Bond\\Angles}$\downarrow$ & Tor.$\downarrow$ & \makecell{Median\\$\Delta E_{relax}$}$\downarrow$ & \makecell{Mean\\$\Delta E_{relax}$}$\downarrow$  &\makecell{Median\\RMSD}$\downarrow$ &\makecell{Mean\\RMSD}$\downarrow$  \\
        \hline
        EQGAT &500 & 3468 & 0.899 & 0.834 & 1.00& 1.15& 8.58& 6.40& 11.1  &0.915&0.975\\
        JODO &500& 673 & 0.963 & 0.879 & 0.77& 0.83& 6.01& 4.74& 7.04  &-&-\\
        Megalodon-quick &500 & - & 0.944 & 0.900& \textbf{0.66}& 0.71& 5.58& 3.19& 5.76  &-&-\\
        \hline
        FlowMol2 &100 & 126.3 & 0.944& 0.746& 1.30& 1.62& 15.0& 17.9& 24.3  &0.992&1.038\\

        SemlaFlow &100 & 150.3 & 0.969& 0.920& 3.10& 2.06& 6.05& 32.3& 91.0  &0.273&0.358\\
        Megalodon-flow &100 & - & 0.987& 0.948& 2.30& 1.62& 5.58& 20.9& 46.9  &-&-\\
        \hline
 VEDA-S &100 & 164.8 & \textbf{0.995}& \textbf{0.988}& 0.69& \textbf{0.41}&\textbf{ 2.71}& \textbf{1.72}&\textbf{2.93}&\textbf{0.096}& \textbf{0.197}\\
 VEDA-S &\textbf{50} & \textbf{84.1} & 0.968& 0.966& 0.86& 0.60& 5.10& 2.99&8.38&0.236&0.355\\
         \hline
    \end{tabular}
    }
\caption{\textbf{Performance comparison on the GEOM-DRUGS dataset.} Our model, VEDA-S (bottom), is compared against denoising-based (top) and flow-based (middle) methods. 
\textbf{Bold} values indicate the best result. 
Metrics include Molecular Stability (MS), Validity \& Connectivity (V\&C), and Mean Absolute Error (MAE) for bond lengths, angles, and torsions. 
All metrics are lower-is-better except for MS and V\&C.
Time (s) reports the average wall-clock time for generating one molecule. 
NFE is the Number of Function Evaluations. 
RMSD values for some models are missing as they were not in the original benchmark. 
For each model, 5000 molecules were evaluated, the full results with confidence intervals and sampling at other steps are included in \ifincludeappendix
  Appendix~\ref{appendix:supplementary_exp}.
\else
extended version.
\fi}

\label{tab:geom_drugs_open_eval}
\end{table*}

\paragraph{Datasets and Metrics}
We evaluate our models on QM9~\cite{ramakrishnan2014quantum} and GEOM-DRUGS~\cite{axelrod2022geom} using distinct protocols. For QM9, we report standard metrics for chemical correctness: Atom Stability (Atom Sta.), Molecule Stability (Mol Sta.), Validity, and Unique percentage. For the more complex GEOM-DRUGS dataset, our main analysis followed the protocol proposed by \citet{nikitin2025geom}, which prioritizes physical realism, reporting the fixed version of molecular stability and validity alongside key metrics from GFN2-xTB geometry optimization, such as relaxation energy ($\Delta E_{relax}$), post-optimization RMSD, and the structural difference after optimization. A supplementary analysis using the MMFF94~\cite{halgren1996merck} force field to assess conformational energy and strain is detailed in the \ifincludeappendix
Appendix~\ref{appendix:supplementary_exp}.
\else
  extended version.
\fi\noindent 
\begin{figure}[htb]
    \centering
    \includegraphics[width=1\columnwidth]{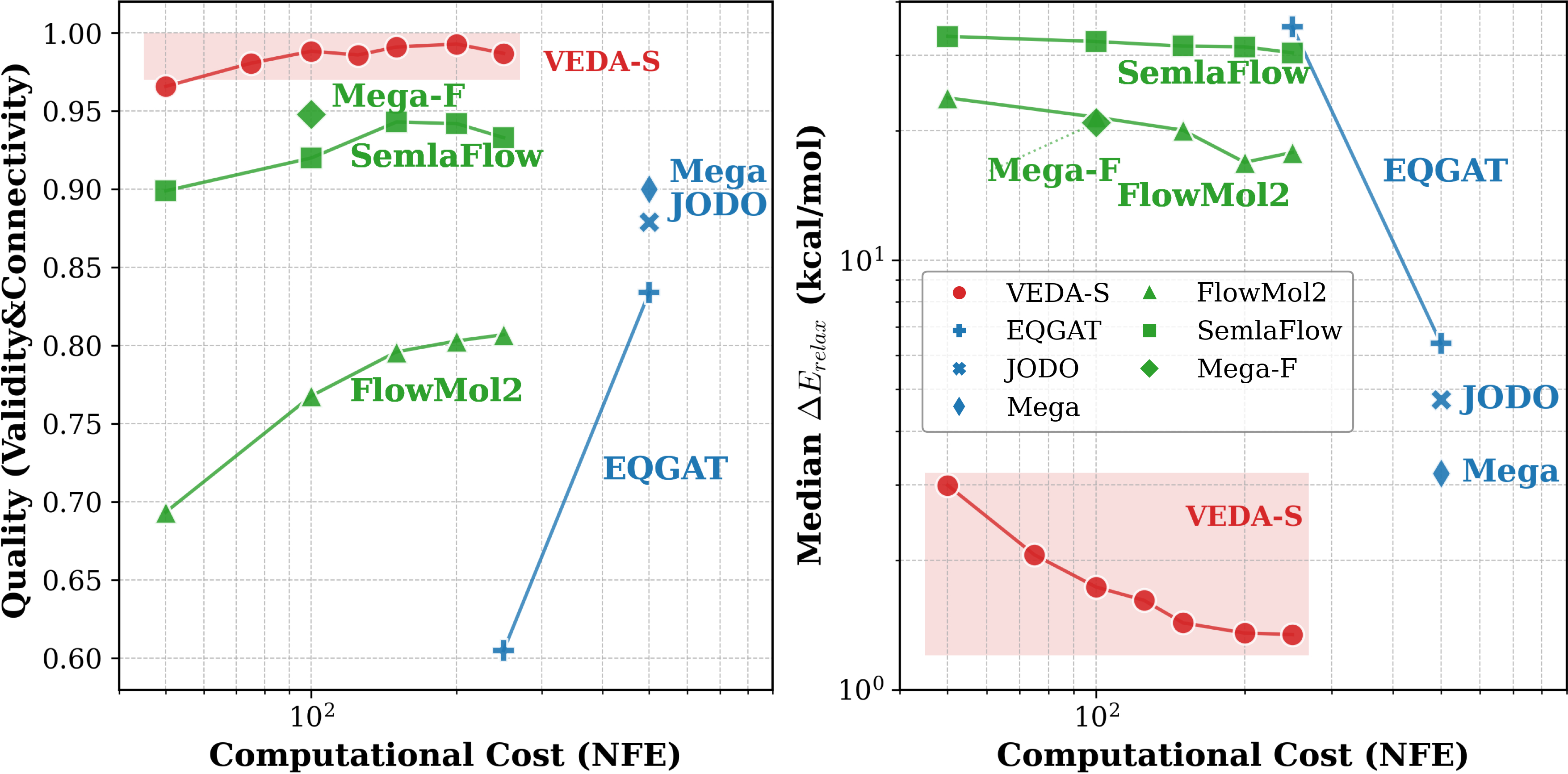}
    \caption{Trade-off between Generation Quality and Computational Cost. The figure compares our model, VEDA-S (red), against flow-based (green) and denoising-based (blue) models. The horizontal axis represents the computational cost, measured by the Number of Function Evaluations (NFE). (Left) Quality measured by molecule Validity and Connectivity, where higher values are better. (Right) Quality measured by the median energy difference ($\Delta E_{relax}$), where lower values are better.}
\label{fig:nfe-quality-trade-off}
\end{figure}

\paragraph{Baselines}
We compare VEDA against leading 3D molecular generative models, grouped by bond treatment. For methods that do not explicitly generate bond structures, we consider EDM~\cite{hoogeboom2022equivariant}, GeoLDM~\cite{xu2023geometric}, EquiFM~\cite{song2023equivariant}, GeoBFN~\cite{song2023unified}, UniGEM~\cite{feng2024unigem} and GOAT~\cite{hongaccelerating_GOAT}, all of which use equivariant architectures. For methods with explicit bond modeling, we compare to SemlaFlow~\cite{irwin2025semlaflow}, FlowMol2~\cite{dunn2024exploring}, and JODO~\cite{huang2023learning}, EQGAT-diff~\cite{lenavigating_EQGAT_diff} and Megalodon~\cite{reidenbach2025applications}.
\noindent
\nopagebreak

\subsection{Main Results}
\paragraph{Results on QM9}
As shown in Table~\ref{tab:uncond_qm9}, our model VEDA achieves state-of-the-art performance on the QM9 dataset. 
In the bond-implicit setting, VEDA-E achieves the highest valid and unique rate (97.9\%) with only 50 NFE, significantly outperforming methods like EDM (90.9\% at 1000 NFE) and GeoLDM (92.6\% at 1000 NFE).
In the bond-explicit setting, VEDA-S achieves nearly perfect atom stability and molecular stability, and a valid and unique score of 98.9\% at just 100 NFE. Even at 30 NFE, it remains competitive with most baselines, highlighting its efficiency.

\begin{figure*}[htp]
    \centering
    \includegraphics[width=\linewidth]{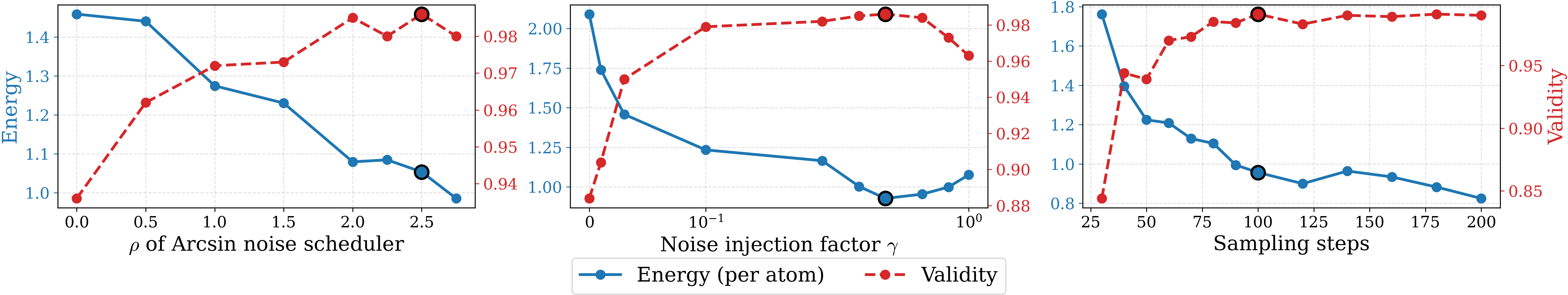}
    \caption{
    Ablation study of key hyperparameters on GEOM-DRUGS: We report MMFF94 energy and validity when varying (left) arcsin noise factor $\rho$, (middle) noise injection level $\gamma$, and (right) sampling steps. Black circles indicate selected values ($\rho{=}2.5$, $\gamma{=}0.4$, 100 steps).
    }
\label{fig:hyper_selection}
\end{figure*}
\paragraph{Results on GEOM-DRUGS}
We evaluate our model on the GEOM-DRUGS dataset using the protocol from \citet{nikitin2025geom}, with detailed metrics provided in Table~\ref{tab:geom_drugs_open_eval}. As the results show, our VEDA-S model achieves a new state-of-the-art performance across nearly all metrics. Notably, this is accomplished in just 100 function evaluations (NFE), significantly outperforming computationally expensive 500-step models like Megalodon, and surpassing previous flow-based approaches by a large margin in geometric accuracy.

This superior performance is not just in final quality but also in the trade-off between sampling efficiency and generation quality, as shown in Figure~\ref{fig:nfe-quality-trade-off}. The figure plots generation quality as a function of NFE. Quality is assessed by both validity and geometric stability, measured via median relaxation energy. The results are striking: VEDA-S (in red) consistently occupies the optimal region in both plots. It achieves the high efficiency of flow-based models (green) while delivering a geometric accuracy (lower $\Delta E_{relax}$) that is an order of magnitude better than all competitors.

To further validate this comparison, we also tested noise‑injected sampling for SemlaFlow \ifincludeappendix
(See Appendix~\ref{appendix:supplementary_exp_SemlaFlow})
\else
\fi, but observed no significant gains. However, this modification did not lead to significant improvements. This finding suggests that VEDA's superior performance arises not merely from the presence of noise, but from its systematic framework design. Even on saturated metrics like molecular stability, our model pushes the scores closer to 100\%. Furthermore, even with just 50 steps, VEDA-S remains highly competitive, demonstrating its excellent trade-off between sampling efficiency and generation quality.

\nopagebreak
\noindent
\subsection{Ablation Studies}
\paragraph{Component Ablation}
Table~\ref{tab:ablation_geom} presents an ablation study on the GEOM-DRUGS dataset. 
We start from the full VEDA model and progressively remove key components to assess their impact. Disabling preconditioning refers to setting $\alpha_t$ to be 0. 
It results in a small drop in validity and modest increases in RMSD and relaxation error, showing the stabilizing effect of preconditioning. 
Removing the noise injection leads to a substantial drop in all metrics, highlighting its importance in guiding the denoising process. 
Introducing the optimal transport (OT) alignment mechanism, as used in SemlaFlow~\cite{irwin2025semlaflow}, increases validity but significantly worsens energy and RMSD. 
We attribute this to OT alignment violating the independence assumption between noise and original molecular coordinates, which causes the denoising direction to become overly correlated with the aligned noise. 
This results in a collapse of coordinate variance during sampling and degrades overall sample quality. 
These results show that each module in VEDA provides complementary improvements to accuracy and validity.
\noindent
\begin{table}[htp]
\small
\setlength{\tabcolsep}{1mm}
\centering

\begin{tabular}{l|ccc|cccc}
Variant & OT & Noise & Pre & V\&C ↑& M.S.↑& E. ↓& R. ↓ \\
\hline
VEDA (Full) & \xmark& \cmark& \cmark& 98.8& 99.5& 1.72& 0.096\\
– Pre        & \xmark& \cmark& \xmark& 97.9& 98.5& 2.14& 0.153\\
– Noise      & \xmark& \xmark& \cmark&95.9&97.6&6.29&0.382\\
– Noise \& Pre      & \xmark& \xmark& \xmark& 93.8& 92.9& 10.1& 0.409\\
+ OT Align& \cmark& \xmark& \xmark& 91.7& 96.0& 23.5& 0.557\\
SemlaFlow & \cmark& \xmark& \xmark& 92.0& 96.9& 32.3& 0.273\\
\end{tabular}
\caption{
\textbf{Ablation study on GEOM-DRUGS.} Each variant enables (\cmark) or disables (\xmark) key components of VEDA: OT alignment (OT), noise injection (Noise), and preconditioning (Pre). V\&C ↑ and M.S.↑ are reported as percentage-based metrics. The last two rows share the same component configuration and differ only in their time parameterization (VE vs. FM).}
\label{tab:ablation_geom}
\end{table}
\paragraph{Hyperparameter Sensitivity}
We performed a sensitivity analysis to support our final hyperparameter choices, as shown in Figure~\ref{fig:hyper_selection}. 
For the arcsin scheduler factor, $\rho$, the model performs robustly across $\rho \in [2.0, 2.75]$, approaching the theoretical limit where the scheduler function remains monotonic. 
For the noise injection level $\gamma$ follows the expected annealing behavior, where an intermediate value achieves the best trade-off. 
Finally, validity saturates around 100 sampling steps, offering an efficient trade-off between quality and cost.

\section{Discussion}
\label{sec:discussion}
Our findings regarding optimal transport alignment, used in prior works like SemlaFlow~\cite{irwin2025semlaflow}, reveals a critical design principle: training strategies must not only improve the loss but also consider their impact on downstream sampling. Our ablation study shows that seemingly beneficial optimizations can violate core statistical assumptions—like noise-data independence—that are crucial for generating diverse and valid molecules.

While two-step (e.g., 2D or SMILES-to-3D) generation pipeline are often considered simpler and more scalable, we argue that native 3D generative capability is essential for designing functional molecules in constrained environments like protein pockets. This necessity is underscored by recent benchmarks showing that current target-aware diffusion models frequently produce invalid geometric structures, despite their high docking scores~\cite{baillif2024benchmarking}. Therefore, enhancing the foundational 3D generative model, as we have focused on in this work, is critical for future success in structure-based design.
\nopagebreak
\section{Conclusions}
In this work, we introduced \textbf{VEDA}, a novel equivariant generative model for 3D molecular structures that unifies continuous and discrete generative processes within a single framework. Through comprehensive experiments on QM9 and GEOM-DRUGS, we show that VEDA not only surpasses strong baselines like GeoBFN and SemlaFlow  in structural validity and stability metrics, but also delivers efficient generation thanks to its unified diffusion framework. Ablations studies validate that both the amplified VE noise schedule and the discrete sampling mechanism are critical to VEDA’s performance.

The success of VEDA in unconditional generation establishes a robust foundation, opening a clear and immediate path toward property-guided molecular design.  A natural extension is to condition the generative process on target chemical properties such as binding affinity or solubility, leveraging VEDA's unified framework to guide generation toward desired molecular profiles. Beyond conditional generation, we identify a fundamental limitation shared across current diffusion models: reliance on implicit velocity prediction through score matching. Future equivariant architectures should be designed to explicitly output the time-dependent vector field $v_\theta(\mathbf{x}, t)$. This would align training objectives with advanced samplers like Consistency Flow Matching~\cite{lusimplifying} and MeanFlow~\cite{geng2025mean}, further improving sampling efficiency and unlocking one- or few-step integration.
 
The power of VEDA lies in its core approach: it uses VE diffusion to smooth the geometric landscape with controlled noise, while seamlessly integrating discrete atomic features. This principled fusion of dynamics and geometry offers an effective blueprint for the future of molecular design.

\bibliography{aaai2026}

@article{ho2020denoising,
  title={Denoising diffusion probabilistic models},
  author={Ho, Jonathan and Jain, Ajay and Abbeel, Pieter},
  journal={Advances in neural information processing systems},
  volume={33},
  pages={6840--6851},
  year={2020}
}

@inproceedings{songscore,
  title={Score-Based Generative Modeling through Stochastic Differential Equations},
  author={Song, Yang and Sohl-Dickstein, Jascha and Kingma, Diederik P and Kumar, Abhishek and Ermon, Stefano and Poole, Ben},
  booktitle={International Conference on Learning Representations},
  year=2021
}

@article{karras2022elucidating,
  title={Elucidating the design space of diffusion-based generative models},
  author={Karras, Tero and Aittala, Miika and Aila, Timo and Laine, Samuli},
  journal={Advances in neural information processing systems},
  volume={35},
  pages={26565--26577},
  year={2022}
}

@article{gao2024diffusion,
  title={Diffusion meets flow matching: Two sides of the same coin},
  author={Gao, Ruiqi and Hoogeboom, Emiel and Heek, Jonathan and Bortoli, VD and Murphy, Kevin P and Salimans, Tim},
  journal={The Internet},
  year={2024}
}

@inproceedings{hoogeboom2022equivariant,
  title={Equivariant diffusion for molecule generation in 3d},
  author={Hoogeboom, Emiel and Satorras, V{\i}ctor Garcia and Vignac, Cl{\'e}ment and Welling, Max},
  booktitle={International conference on machine learning},
  pages={8867--8887},
  year={2022},
  organization={PMLR}
}

@inproceedings{liaoequiformer,
  title={Equiformer: Equivariant Graph Attention Transformer for 3D Atomistic Graphs},
  author={Liao, Yi-Lun and Smidt, Tess},
  booktitle={The Eleventh International Conference on Learning Representations},
  year={2023}
}

@article{zhang2025d3mes,
  title={D3MES: Diffusion Transformer with multihead equivariant self-attention for 3D molecule generation},
  author={Zhang, Zhejun and Chen, Yuanping and Chu, Shibing},
  journal={arXiv preprint arXiv:2501.07077},
  year={2025}
}

@inproceedings{xu2023geometric,
  title={Geometric latent diffusion models for 3d molecule generation},
  author={Xu, Minkai and Powers, Alexander S and Dror, Ron O and Ermon, Stefano and Leskovec, Jure},
  booktitle={International Conference on Machine Learning},
  pages={38592--38610},
  year={2023},
  organization={PMLR}
}

@article{song2023equivariantFM,
  title={Equivariant flow matching with hybrid probability transport for 3d molecule generation},
  author={Song, Yuxuan and Gong, Jingjing and Xu, Minkai and Cao, Ziyao and Lan, Yanyan and Ermon, Stefano and Zhou, Hao and Ma, Wei-Ying},
  journal={Advances in Neural Information Processing Systems},
  volume={36},
  pages={549--568},
  year={2023}
}

@inproceedings{satorras2021n_EGNN,
  title={E (n) equivariant graph neural networks},
  author={Satorras, V{\i}ctor Garcia and Hoogeboom, Emiel and Welling, Max},
  booktitle={International conference on machine learning},
  pages={9323--9332},
  year={2021},
  organization={PMLR}
}

@article{fuchs2020se_Etransformer,
  title={Se (3)-transformers: 3d roto-translation equivariant attention networks},
  author={Fuchs, Fabian and Worrall, Daniel and Fischer, Volker and Welling, Max},
  journal={Advances in neural information processing systems},
  volume={33},
  pages={1970--1981},
  year={2020}
}

@article{abramson2024accurate,
  title={Accurate structure prediction of biomolecular interactions with AlphaFold 3},
  author={Abramson, Josh and Adler, Jonas and Dunger, Jack and Evans, Richard and Green, Tim and Pritzel, Alexander and Ronneberger, Olaf and Willmore, Lindsay and Ballard, Andrew J and Bambrick, Joshua and others},
  journal={Nature},
  pages={1--3},
  year={2024},
  publisher={Nature Publishing Group UK London}
}

@article{axelrod2022geom,
  title={GEOM, energy-annotated molecular conformations for property prediction and molecular generation},
  author={Axelrod, Simon and Gomez, Bombarelli, Rafael},
  journal={Scientific Data},
  volume={9},
  number={1},
  pages={185},
  year={2022},
  publisher={Nature Publishing Group UK London}
}

@article{ramakrishnan2014quantum,
  title={Quantum chemistry structures and properties of 134 kilo molecules},
  author={Ramakrishnan, Raghunathan and Dral, Pavlo O and Rupp, Matthias and Von Lilienfeld, O Anatole},
  journal={Scientific data},
  volume={1},
  number={1},
  pages={1--7},
  year={2014},
  publisher={Nature Publishing Group}
}

@inproceedings{lipman2023flow,
  title={Flow Matching for Generative Modeling},
  author={Lipman, Yaron and Chen, Ricky TQ and Ben-Hamu, Heli and Nickel, Maximilian and Le, Matt},
  booktitle={11th International Conference on Learning Representations, ICLR 2023},
  year={2023}
}

@inproceedings{
irwin2025semlaflow,
title={SemlaFlow -- Efficient 3D Molecular Generation with Latent Attention and Equivariant Flow Matching},
author={Ross Irwin and Alessandro Tibo and Jon Paul Janet and Simon Olsson},
booktitle={The 28th International Conference on Artificial Intelligence and Statistics},
year={2025},
url={https://openreview.net/forum?id=bee2G6pEh0}
}

@article{gat2024discrete,
  title={Discrete flow matching},
  author={Gat, Itai and Remez, Tal and Shaul, Neta and Kreuk, Felix and Chen, Ricky TQ and Synnaeve, Gabriel and Adi, Yossi and Lipman, Yaron},
  journal={Advances in Neural Information Processing Systems},
  volume={37},
  pages={133345--133385},
  year={2024}
}

@inproceedings{campbell2024generative,
  title={Generative Flows on Discrete State-Spaces: Enabling Multimodal Flows with Applications to Protein Co-Design},
  author={Campbell, Andrew and Yim, Jason and Barzilay, Regina and Rainforth, Tom and Jaakkola, Tommi},
  booktitle={International Conference on Machine Learning},
  pages={5453--5512},
  year={2024},
  organization={PMLR}
}

@article{miao2015gaussian,
  title={Gaussian accelerated molecular dynamics: unconstrained enhanced sampling and free energy calculation},
  author={Miao, Yinglong and Feher, Victoria A and McCammon, J Andrew},
  journal={Journal of chemical theory and computation},
  volume={11},
  number={8},
  pages={3584--3595},
  year={2015},
  publisher={ACS Publications}
}

@inproceedings{hongaccelerating_GOAT,
  title={Accelerating 3D Molecule Generation via Jointly Geometric Optimal Transport},
  author={Hong, Haokai and Lin, Wanyu and Tan, Kay Chen},
  booktitle={The Thirteenth International Conference on Learning Representations},
  year={2024}
}

@article{song2023equivariant,
  title={Equivariant flow matching with hybrid probability transport for 3d molecule generation},
  author={Song, Yuxuan and Gong, Jingjing and Xu, Minkai and Cao, Ziyao and Lan, Yanyan and Ermon, Stefano and Zhou, Hao and Ma, Wei-Ying},
  journal={Advances in Neural Information Processing Systems},
  volume={36},
  pages={549--568},
  year={2023}
}

@inproceedings{song2023unified,
  title={Unified generative modeling of 3d molecules with bayesian flow networks},
  author={Song, Yuxuan and Gong, Jingjing and Zhou, Hao and Zheng, Mingyue and Liu, Jingjing and Ma, Wei-Ying},
  booktitle={The Twelfth International Conference on Learning Representations},
  year={2023}
}

@inproceedings{lenavigating_EQGAT_diff,
  title={Navigating the Design Space of Equivariant Diffusion-Based Generative Models for De Novo 3D Molecule Generation},
  author={Le, Tuan and Cremer, Julian and Noe, Frank and Clevert, Djork-Arn{\'e} and Sch{\"u}tt, Kristof T},
  year={2024},
  booktitle={The Twelfth International Conference on Learning Representations}
}

@article{feng2024unigem,
  title={UniGEM: A Unified Approach to Generation and Property Prediction for Molecules},
  author={Feng, Shikun and Ni, Yuyan and Lu, Yan and Ma, Zhi-Ming and Ma, Wei-Ying and Lan, Yanyan},
  journal={arXiv preprint arXiv:2410.10516},
  year={2024}
}

@article{zhang2025unraveling,
  title={Unraveling the potential of diffusion models in small-molecule generation},
  author={Zhang, Peining and Baker, Daniel and Song, Minghu and Bi, Jinbo},
  journal={Drug Discovery Today},
  pages={104413},
  year={2025},
  publisher={Elsevier}
}

@article{dunn2024exploring,
  title={Exploring Discrete Flow Matching for 3D De Novo Molecule Generation},
  author={Dunn, Ian and Koes, David R},
  journal={ArXiv},
  pages={arXiv--2411},
  year={2024}
}

@inproceedings{devlin2019bert,
  title={Bert: Pre-training of deep bidirectional transformers for language understanding},
  author={Devlin, Jacob and Chang, Ming-Wei and Lee, Kenton and Toutanova, Kristina},
  booktitle={Proceedings of the 2019 conference of the North American chapter of the association for computational linguistics: human language technologies, volume 1 (long and short papers)},
  pages={4171--4186},
  year={2019}
}

@inproceedings{chang2022maskgit,
  title={Maskgit: Masked generative image transformer},
  author={Chang, Huiwen and Zhang, Han and Jiang, Lu and Liu, Ce and Freeman, William T},
  booktitle={Proceedings of the IEEE/CVF conference on computer vision and pattern recognition},
  pages={11315--11325},
  year={2022}
}

@article{austin2021structured,
  title={Structured denoising diffusion models in discrete state-spaces},
  author={Austin, Jacob and Johnson, Daniel D and Ho, Jonathan and Tarlow, Daniel and Van Den Berg, Rianne},
  journal={Advances in neural information processing systems},
  volume={34},
  pages={17981--17993},
  year={2021}
}

@inproceedings{nie2025large,
  title={Large Language Diffusion Models},
  author={Nie, Shen and Zhu, Fengqi and You, Zebin and Zhang, Xiaolu and Ou, Jingyang and Hu, Jun and ZHOU, JUN and Lin, Yankai and Wen, Ji-Rong and Li, Chongxuan},
  booktitle={ICLR 2025 Workshop on Deep Generative Model in Machine Learning: Theory, Principle and Efficacy},
    year={2025}
}

@article{nikitin2025geom,
  title={GEOM-Drugs Revisited: Toward More Chemically Accurate Benchmarks for 3D Molecule Generation},
  author={Nikitin, Filipp and Dunn, Ian and Koes, David Ryan and Isayev, Olexandr},
  journal={arXiv preprint arXiv:2505.00169},
  year={2025}
}

@article{geng2025mean,
  title={Mean flows for one-step generative modeling},
  author={Geng, Zhengyang and Deng, Mingyang and Bai, Xingjian and Kolter, J Zico and He, Kaiming},
  journal={arXiv preprint arXiv:2505.13447},
  year={2025}
}

@inproceedings{he2016deep,
  title={Deep residual learning for image recognition},
  author={He, Kaiming and Zhang, Xiangyu and Ren, Shaoqing and Sun, Jian},
  booktitle={Proceedings of the IEEE conference on computer vision and pattern recognition},
  pages={770--778},
  year={2016}
}

@article{halgren1996merck,
  title={Merck molecular force field. I. Basis, form, scope, parameterization, and performance of MMFF94},
  author={Halgren, Thomas A},
  journal={Journal of computational chemistry},
  volume={17},
  number={5-6},
  pages={490--519},
  year={1996},
  publisher={Wiley Online Library}
}

@article{reidenbach2025applications,
  title={Applications of Modular Co-Design for De Novo 3D Molecule Generation},
  author={Reidenbach, Danny and Nikitin, Filipp and Isayev, Olexandr and Paliwal, Saee},
  journal={arXiv preprint arXiv:2505.18392},
  year={2025}
}

@inproceedings{kimtrain,
  title={Train for the Worst, Plan for the Best: Understanding Token Ordering in Masked Diffusions},
  author={Kim, Jaeyeon and Shah, Kulin and Kontonis, Vasilis and Kakade, Sham M and Chen, Sitan},
  booktitle={Forty-second International Conference on Machine Learning},
  year={2025}
}

@article{huang2023learning,
  title={Learning Joint 2D \& 3D Diffusion Models for Complete Molecule Generation},
  author={Huang, Han and Sun, Leilei and Du, Bowen and Lv, Weifeng},
  journal={CoRR},
  year={2023}
}

@article{baillif2024benchmarking,
  title={Benchmarking structure-based three-dimensional molecular generative models using GenBench3D: ligand conformation quality matters},
  author={Baillif, Benoit and Cole, Jason and McCabe, Patrick and Bender, Andreas},
  journal={arXiv preprint arXiv:2407.04424},
  year={2024}
}

@article{watson2023novo,
  title={De novo design of protein structure and function with RFdiffusion},
  author={Watson, Joseph L and Juergens, David and Bennett, Nathaniel R and Trippe, Brian L and Yim, Jason and Eisenach, Helen E and Ahern, Woody and Borst, Andrew J and Ragotte, Robert J and Milles, Lukas F and others},
  journal={Nature},
  volume={620},
  number={7976},
  pages={1089--1100},
  year={2023},
  publisher={Nature Publishing Group UK London}
}

@inproceedings{lusimplifying,
  title={Simplifying, Stabilizing and Scaling Continuous-time Consistency Models},
  author={Lu, Cheng and Song, Yang},
  booktitle={The Thirteenth International Conference on Learning Representations},
  year={2025}
}

@article{jing2022torsional,
  title={Torsional diffusion for molecular conformer generation},
  author={Jing, Bowen and Corso, Gabriele and Chang, Jeffrey and Barzilay, Regina and Jaakkola, Tommi},
  journal={Advances in neural information processing systems},
  volume={35},
  pages={24240--24253},
  year={2022}
}

@inproceedings{huang2023mdm,
  title={Mdm: Molecular diffusion model for 3d molecule generation},
  author={Huang, Lei and Zhang, Hengtong and Xu, Tingyang and Wong, Ka-Chun},
  booktitle={Proceedings of the AAAI Conference on Artificial Intelligence},
  volume={37},
  number={4},
  pages={5105--5112},
  year={2023}
}
\ifincludeappendix
\appendix
\setcounter{secnumdepth}{1}
\section{Experiment Details}
\label{appendix: exp_details}
This section describes the full experiment setup in this paper.
\paragraph{Hyperparameters}
All experiments are conducted on a single NVIDIA A100 40GB GPU. The search space for model and training hyperparameters are listed in Table \ref{tab:hyperparameters}. VEDA-S was trained using the data splits and adaptive data loader from \citet{irwin2025semlaflow}.
\begin{table*}
    \setlength{\tabcolsep}{10pt}
    \centering
        \begin{tabular}{l|cc|cc}
            Hyperparameter & \multicolumn{2}{c|}{QM9} & \multicolumn{2}{c}{GEOM-DRUGS} \\
            & VEDA-E & VEDA-S & VEDA-E & VEDA-S\\
            \hline
            Learning rate & 3e-4 & 3e-4 & 1e-3 & 3e-4\\
            Warm up steps & 2000 & 15000 & 2000 & 10000\\
            Batch Size & 32 & N/A & 4 & N/A\\
            Batch Cost & N/A & 4096 & N/A & 4096\\
            EMA decay rate & 0.9999 & 0.999 & 0.9999 & 0.999\\
            $T_{min}$ & 0.001 & 0.001 & 0.001 & 0.001\\
            $T_{max}$ & 80 & 80 & 80 & 80\\
            $\sigma_d$& 1& 1& 1& 1\\
            Training iterations & 2000 & 300 & 10 & 200\\
            
            Optimizer& Adam& Adam & Adam &Adam \\
            EGNN layer& 9& N/A & 4 & N/A \\
            Features per layer& 256&256 &256&256 \\
            Activations Function& SiLU & SiLU & SiLU & SiLU\\
            Semla $n_{head}$& N/A& 32 & N/A & 32 \\
            Loss weight ($\lambda_x, \lambda_a, \lambda_b, \lambda_c$) & (1.0, 1.0, N/A, 1.0)& (1.0, 0.2, 0.5, 1.0) & (1.0, 1.0, N/A, 1.0) & (1.0, 0.2, 1.0, 1.0) \\
            Noise scheduler $\rho$ & 2.5& 2.5 & 2.5 & 2.5 \\
            categorical noise level $\eta$ & N/A&1 & N/A & 1\\
            Sampling Temperature ($\tau$) & 1.0 & 0.9 & 1.0 & 1.0 \\
            Max $w_{\text{cat}}$ (Clip) & N/A & 10 & N/A & 10 \\
            Prediction Mode & Constant & Constant & Constant & Adaptive \\
            Cat. Loss Weighting & N/A & $1/m(t)$ & N/A & $1/m(t)$ \\

        \end{tabular}
        \caption{Hyperparameters used for training VEDA-E and VEDA-S models}\label{tab:hyperparameters}
\end{table*}


\paragraph{Metrics}
To enable fair and consistent evaluation, we follow the metrics and baseline settings widely adopted in previous works on 3D molecular generation (e.g., EDM\cite{hoogeboom2022equivariant}, SemlaFlow\cite{irwin2025semlaflow}) and the benchmarks~\cite{nikitin2025geom}.

\begin{itemize}
    \item \textbf{Atom Stability}: Percentage of atoms that satisfy standard valence rules.
    \item \textbf{Molecule Stability}: Percentage of molecules in which all atoms are stable.
    \item \textbf{Validity}: Percentage of molecules that can be parsed into valid SMILES strings via RDKit.
    \item \textbf{Validity \& Uniqueness}: Fraction of unique samples within the set of valid molecules.
    \item \textbf{Conformational Energy (MMFF)}: Estimated using the MMFF94 force field in RDKit; lower energy implies more realistic molecular structures.
    \item \textbf{Strain Energy (MMFF)}: Energy difference between the initial and optimized conformations under MMFF; reflects post-processing effort for physical plausibility.
    \item \textbf{RMSD (xTB)}: Root Mean Square Deviation between the initial and optimized conformations using the xTB toolkit.
    \item \textbf{RMSD (MMFF)}: RMSD computed under MMFF force field.
    \item $\Delta E_{\text{relax}}$(xTB): Energy decrease after xTB geometry optimization.
    \item \textbf{Bond Length Deviation}: Mean absolute error (MAE) of bond lengths between initial and optimized structures: 
    \[
    \Delta r_{ij} = r^{\text{init}}_{ij} - r^{\text{opt}}_{ij}
    \]
    \item \textbf{Bond Angle Deviation}: MAE between initial and optimized bond angles:
    \[
    \Delta \theta_{ijk} = \min\left( \left|\theta^{\text{init}}_{ijk} - \theta^{\text{opt}}_{ijk} \right|,\ 180^\circ - \left|\theta^{\text{init}}_{ijk} - \theta^{\text{opt}}_{ijk} \right| \right)
    \]
    \item \textbf{Torsion Angle Deviation}: MAE between initial and optimized torsion angles (a.k.a. dihedral angles):
    \[
    \Delta \phi_{ijkl} = \min\left( \left|\phi^{\text{init}}_{ijkl} - \phi^{\text{opt}}_{ijkl} \right|,\ 360^\circ - \left|\phi^{\text{init}}_{ijkl} - \phi^{\text{opt}}_{ijkl} \right| \right)
    \]
\end{itemize}

\paragraph{Self-Conditioning}
Following the configuration of the SemlaFlow baseline \cite{irwin2025semlaflow}, we incorporate a self-conditioning mechanism for the \textbf{VEDA-S} model to enhance generation quality. This technique involves augmenting the model input with a preliminary estimate of the clean data, denoted as $\tilde{x}_0$.
\begin{itemize}
    \item \textbf{Training:} We employ a stochastic conditioning strategy. With a probability of $0.5$, the model performs an initial forward pass to estimate $\tilde{x}_0$, which is then concatenated with the noisy input for the final prediction. In the remaining $50\%$ of cases, the conditioning input is set to zero (null condition).
    \item \textbf{Sampling:} Self-conditioning is active for all inference steps except the initial one (where no prior estimate exists). The estimate of the clean data derived from the previous timestep $t_{i-1}$ is used as the condition for the current step $t_i$.
\end{itemize}
Note that this mechanism is exclusive to VEDA-S and is not employed in the VEDA-E architecture.

\section{Supplementary experiments}
\label{appendix:supplementary_exp}
\begin{table*}[htp]
\centering
\begin{tabular}{l|c|ccc|ccc}
\hline
Methods& NFE & \makecell{Atom\\Stab} $\uparrow$& \makecell{Mol\\Stab} $\uparrow$ & Valid $\uparrow$ & $\text{Energy}_{\text{MMFF}}$ $\downarrow$& $\text{Strain}_{\text{MMFF}}$ $\downarrow$  & $\text{RMSD}_{\text{MMFF}}$ $\downarrow$ \\
\hline
EDM & 1000 & 81.3 & N/A & 92.6 &- & - & -  \\
GeoLDM & 1000 & 84.4 & N/A & \textbf{99.6}&- & - & -  \\
 UniGEM& 1000& 85.1& N/A & 98.4& & &\\
EquiFM & 1000 & 84.1 & N/A &98.9 &- & - & -  \\
GOAT & 90 &84.8 & N/A & 96.2 &- & - & -  \\
VEDA-E & 50 & 88.7& 3.6& 97.3& & &\\
VEDA-E & 100 &\textbf{90.0} &\textbf{8.2} &99.5&- & - & -  \\
\hline
MiDi     & 500 & 99.8  & 91.6  & 77.8    &- & - & -  \\
EQGAT-diff$^*$        & 500  & 99.8  & 93.4 & 94.6    & 68.8 & 28.9 & 1.115\\
SemlaFlow      & 100    & 99.8  & 97.7 & 95.2  & 127.5 & 88.9 & 0.857\\
FlowMol-CTMC$^*$    & 250 &    99.9  &95.7     &90.1    & 71.1     &    37.5    & 1.255\\
VEDA-S & 100 & \textbf{99.8}& \textbf{98.1}& \textbf{98.9}& \textbf{43.0}& \textbf{12.7}& \textbf{0.554}\\
VEDA-S & 50 & 99.5& 93.3& 95.3& 61.2& 20.7&0.714\\
Data & - & 100 & 100 & 100 & 50.3 & 15.9 & -\\
\hline
\end{tabular}
\caption{
Molecular generation results on the GEOM-DRUGS dataset.
VEDA-S with Semla as the backbone. Molecule stability and validity are reported as percentages, while energy and strain energy are expressed in kcal·mol$^{-1}$. Results marked with $^*$ were reproduced in our own experiments. All results are calculated based on 1000 randomly sampled molecules.
}\label{tab:unconditional_geomDrugs}
\end{table*}

We also report results following the same metrics as SemlaFlow\cite{irwin2025semlaflow}, as shown in Table \ref{tab:unconditional_geomDrugs}. Due to scalability limitations, UniGEM\cite{feng2024unigem} and GeoBFN\cite{song2023unified} are excluded. VEDA maintains high validity and significantly improves energy metrics in both bond-implicit and bond-explicit settings, reinforcing its robustness on large and chemically diverse molecules.

\subsection{Sampling Efficiency and Speed of VEDA‑E on QM9}
We assess VEDA‑E using two samplers—a first‑order method and a second‑order method—and compare against EDM~\cite{hoogeboom2022equivariant} run with both the DDIM and the original DDPM samplers. Figure~\ref{fig:quality_nfe} presents the trade‑off between sample quality and the number of function evaluations (NFE).
\begin{figure*}[htb]
    \centering
    \begin{subfigure}[t]{0.32\linewidth}
        \centering
        \includegraphics[width=\linewidth]{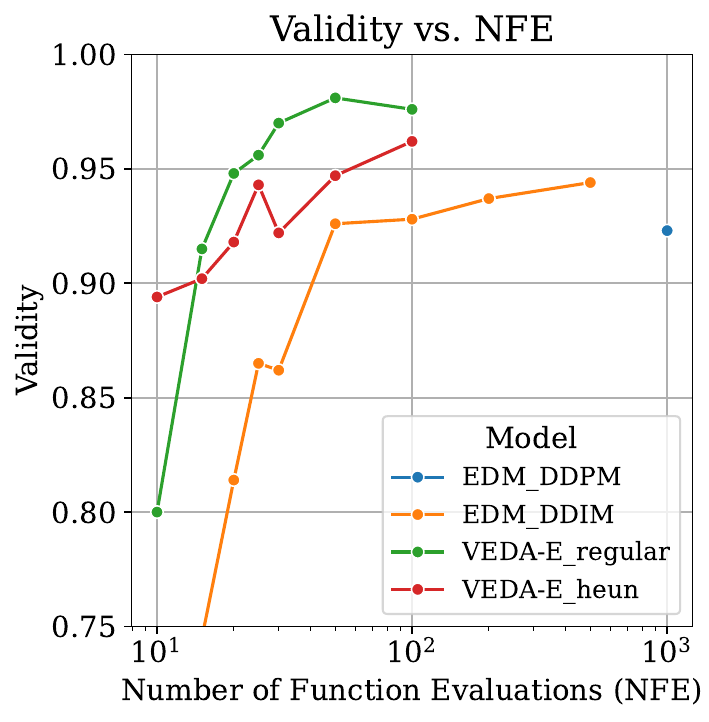}
        \label{fig:validity_nfe}
    \end{subfigure}
    \begin{subfigure}[t]{0.32\linewidth}
        \centering
        \includegraphics[width=\linewidth]{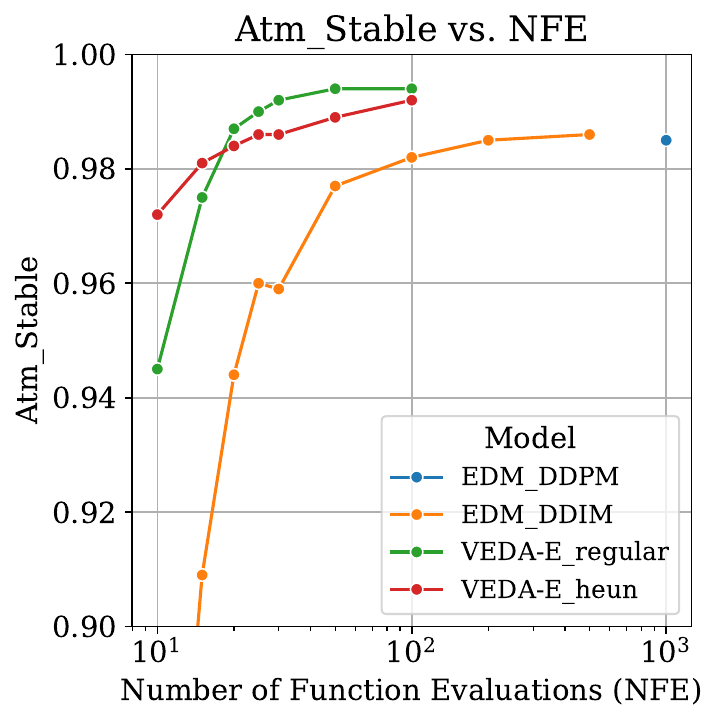}
        \label{fig:atm_stable_nfe}
    \end{subfigure}
    \begin{subfigure}[t]{0.32\linewidth}
        \centering
        \includegraphics[width=\linewidth]{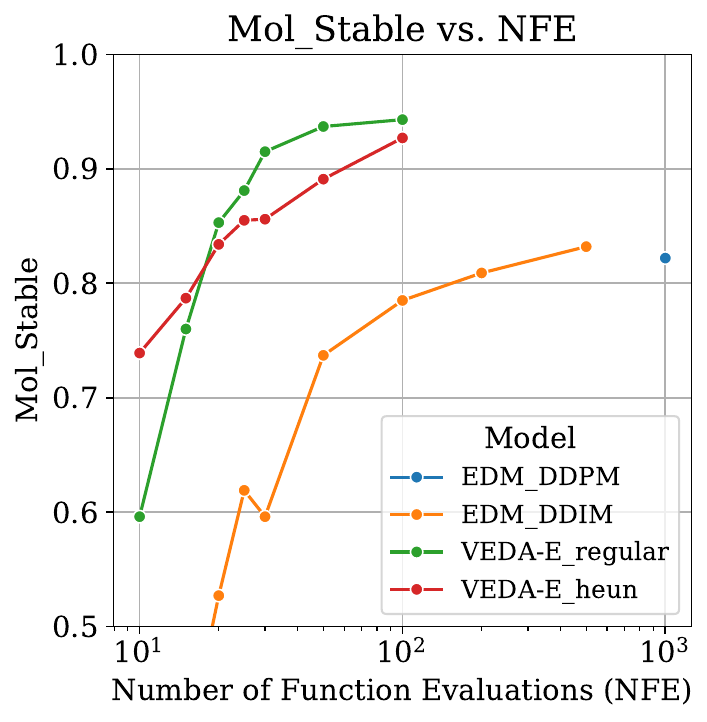}
        \label{fig:mol_stable_nfe}
    \end{subfigure}
    \caption{
    Quality–efficiency trade‑off for VEDA‑E on QM9 under different sampling strategies, compared with EDM~\cite{hoogeboom2022equivariant}. Even at low NFE, our model sustains high validity and structural stability.}
    \label{fig:quality_nfe}
\end{figure*}

\subsection{SemlaFlow on Different Noise Injection}
\label{appendix:supplementary_exp_SemlaFlow}
We evaluate the performance of SemlaFlow under varying levels of noise injection. Specifically, at each noise level $\gamma$, the sampling step adds scaled Gaussian noise $\gamma\epsilon$, where $\epsilon \sim \mathcal{N}(0, 1)$. The results are summarized in Table~\ref{tab:geom_drug_semla_xtb}, using the benchmark protocol from \citet{nikitin2025geom}. 
Although SemlaFlow~\cite{irwin2025semlaflow} reports $\gamma = 0.2$ in their paper, we find that their open-source implementation sets $\gamma = 0$. We therefore include a range of $\gamma$ values in our experiments to assess its effect.
Overall, SemlaFlow does not appear to benefit significantly from increased noise levels. Most metrics, including RMSD and energy relaxation, show marginal or inconsistent changes across different $\gamma$ values, suggesting limited robustness or gain from additional noise injection.


\begin{table*}[tbp]
\centering
\renewcommand{\arraystretch}{0.95}
\small
    
    \begin{tabular}{l|c|cc|ccc|ccll}
        \hline
        Model & NFE& MS$\uparrow$ & \makecell{V\&C}$\uparrow$ & \makecell{Bond\\Length\\($\times10^{-2}$)}$\downarrow$ & \makecell{Bond\\Angles}$\downarrow$ & Torsions$\downarrow$ & \makecell{Median\\$\Delta E_{relax}$}$\downarrow$ & \makecell{Mean\\$\Delta E_{relax}$}$\downarrow$  &\makecell{Median\\RMSD}$\downarrow$ &\makecell{Mean\\RMSD}$\downarrow$  \\
        \hline
SemlaFlow($\gamma$=0)&100 & 0.969& 0.920& 3.10& 2.06& 6.05& 32.3& 91.0  &0.273&0.358\\
SemlaFlow($\gamma$=0.01)& 100 & 0.969& 0.928& 2.90& 1.92& 5.47& 31.7& 72.6& 0.252&0.368\\
SemlaFlow($\gamma$=0.03)& 100 & 0.973& 0.934& 2.88& 1.93& 5.66& 31.9& 67.5& 0.250&0.382\\
SemlaFlow($\gamma$=0.1)& 100 & 0.978& 0.933& 2.87& 1.92& 5.56& 31.6& 67.0& 0.269&0.383\\
SemlaFlow($\gamma$=0.2)& 100 & 0.975& 0.932& 2.93& 1.95& 5.68& 33.8& 77.6& 0.254&0.384\\
SemlaFlow($\gamma$=0.4)& 100 & 0.970& 0.926& 2.88& 1.93& 5.61& 31.6& 71.0& 0.256&0.387\\
    \end{tabular}
\caption{
Performance comparison on the GEOM-DRUGS dataset using the benchmark from \citet{nikitin2025geom}. 
Noise is added according to $v = v + \epsilon \cdot \gamma$, where $\epsilon \sim \mathcal{N}(0, 1)$. 
Although SemlaFlow~\cite{irwin2025semlaflow} reports $\gamma = 0.2$ in their paper, their open-source implementation uses $\gamma = 0$.
}
\label{tab:geom_drug_semla_xtb}
\end{table*}

\begin{figure*}[ht]
    \centering
    \includegraphics[width=1\linewidth]{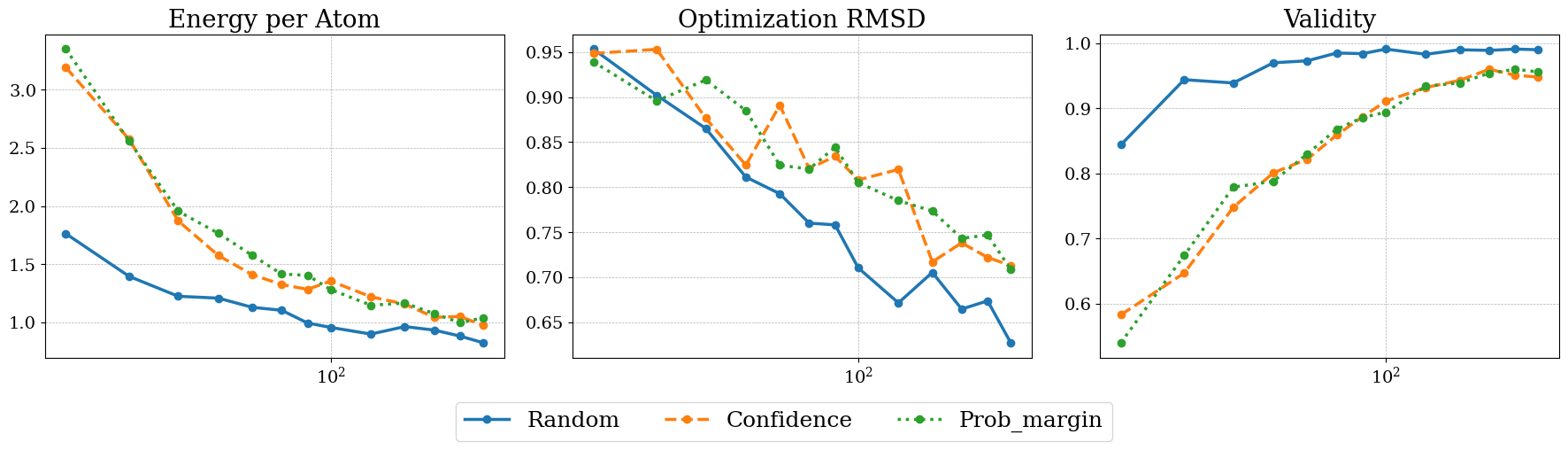}
    \caption{Comparison of re-masking strategies. The x-axis is the sampling steps. The performance of a Uniform Random strategy is compared against two confidence-based heuristics: selecting tokens with the lowest prediction confidence (\texttt{Confidence}) or the smallest probability margin (\texttt{Prob\_margin}). The random strategy consistently achieves lower energy, lower optimization RMSD, and higher validity, indicating its superiority for this generative task.}
    \label{fig:low_conf_remask_hyper_selection}
\end{figure*}
\subsection{Low-Confidence Re-masking Strategies}
\label{appendix:low_conf_remask}
In iterative generative models, a key design choice is the strategy for selecting which tokens to mask and regenerate at each step. An intuitive heuristic, inspired by work like MaskGIT~\cite{chang2022maskgit}, is to focus on regions of low model confidence, with the goal of correcting the most likely errors. To investigate this, we explored two distinct confidence-based re-masking strategies:

\textbf{Lowest Confidence:} This strategy prioritizes re-masking tokens for which the model assigned the lowest probability to the predicted token. This directly targets the model's points of highest uncertainty.

\textbf{Smallest Probability Margin:} This strategy selects tokens where the difference between the probabilities of the most likely and second-most likely predictions is minimal. This targets tokens where the model is highly ambivalent between multiple outcomes.

We compared these two heuristic approaches against a baseline \textbf{Uniform Random} strategy, which selects tokens for re-masking uniformly at random from all available positions.

The results of this comparison are presented in Figure~\ref{fig:low_conf_remask_hyper_selection}, evaluated across three critical metrics: \textit{Energy per Atom}, a measure of the thermodynamic stability of the generated molecule, where lower values indicate more physically plausible structures; \textit{Optimization RMSD (Root Mean Square Deviation)}, the geometric deviation between the as-generated structure and its energy-minimized (i.e., optimized) conformation, where lower values signify a better initial prediction; and \textit{Validity}, the percentage of generated outputs that are chemically valid molecules.

As demonstrated in Figure~\ref{fig:low_conf_remask_hyper_selection}, the Uniform Random strategy consistently and significantly outperforms both confidence-based methods across all three metrics. It converges to states with lower energy, produces geometries with lower RMSD to their optimized counterparts, and achieves a higher validity rate more quickly.

This finding suggests that, for this task, the stochasticity introduced by random selection is more beneficial for exploring the solution space and avoiding local minima than a greedy focus on correcting low-confidence predictions. While seemingly counter-intuitive, this indicates that the model's confidence scores may not be the most reliable guide for iterative refinement.

\section{Ablation Studies}
\subsection{Ablation Study on various Discrete Sampling methods}
\label{appendix:ablation_discrete_sampling}
The choice of the discrete sampling strategy and its associated hyperparameters is critical to the quality of the final generated structures. In Figure \ref{fig:cat_noise_level_hyper_selection}, we present a detailed comparison of the three sampler configurations discussed in the main text. The x-axis represents the hyperparameter $\eta$, which controls the categorical noise level, plotted on a symmetrical log scale to provide clarity for both small and large values. The y-axes correspond to three critical performance metrics: Energy per Atom (lower is better), Optimization RMSD (lower is better), and Validity (higher is better). 
Our analysis reveals that both the fixed and variable $\eta_t$ schedules of \citet{campbell2024generative}'DFM can both achieve top-tier performance, as long as on the appropriate selection of the noise hyperparameter $\eta$. For instance, the fixed $\eta_t$ model (solid orange) reaches its optimal Validity near $\eta$=0.3. Similarly, the variable $\eta_t$ model (dashed blue) finds its performance sweet spot in the region of $\eta$near 1.0. This indicates that the primary factor for success is tuning $\eta$ correctly, rather than the choice between a fixed or variable schedule itself. While both can reach a similar performance minimum, the variable $\eta_t$ schedule appears to offer a slightly wider range of near-optimal performance, potentially making it more robust. Conversely, the fixed $\eta_t$ strategy is simpler and equally effective when $\eta$ is precisely tuned.
  Based on this analysis, we conclude that the performance difference between the fixed and variable $\eta_t$
  schedules is not decisive. Given its slightly broader optimal range, we selected the \citet{campbell2024generative}'DFM sampler with the variable $\eta_t$ schedule for our main experiments, with $\eta$ tuned to 1.0.
\begin{figure*}[htb]
    \centering
    \includegraphics[width=1\linewidth]{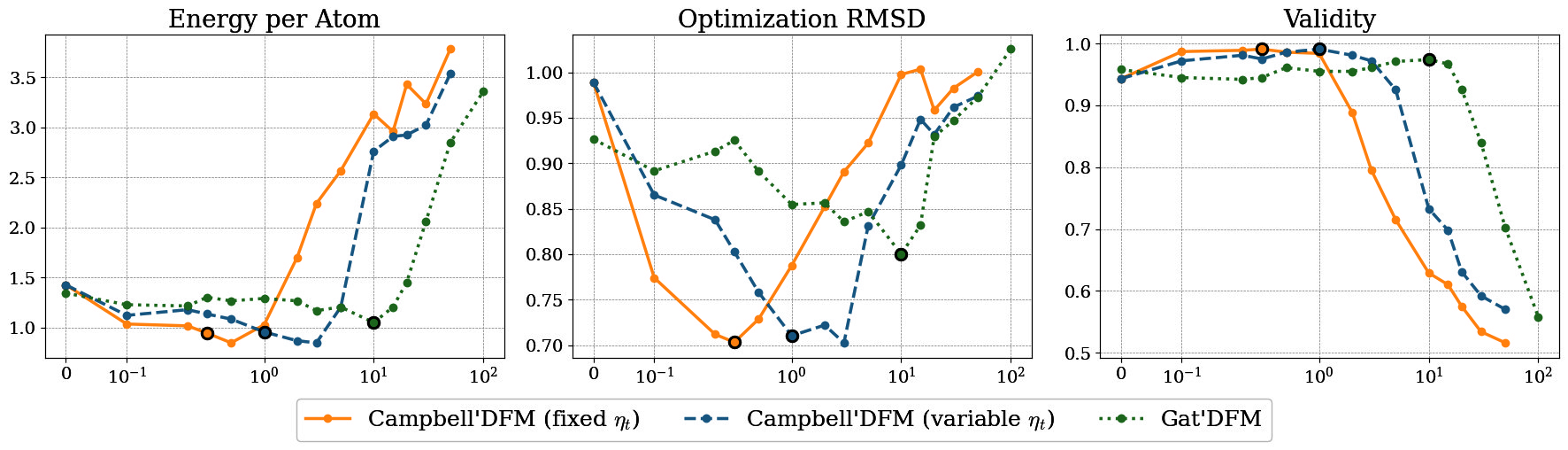}
    \caption{Hyperparameter tuning for DFM sampling strategies. We compare the performance of two distinct sampling approaches derived from the Discrete Flow Matching (DFM) framework. The first approach, based on the transition rate formulation of \citet{campbell2024generative}, is evaluated with both a fixed $\eta_t$(solid orange line) and a variable $\eta_t$ schedule (dashed blue line). The second approach is the simplified velocity-field method from \citet{gat2024discrete}, denoted as Gat'DFM (dotted green line). Performance is measured across three metrics—Energy per Atom, Optimization RMSD, and Validity—as a function of the categorical noise level hyperparameter $\eta$. The results show that the variable $\eta_t$ schedule provides the most robust trade-off, significantly minimizing optimization RMSD while maintaining high structural validity across a wide range of noise levels.}
  
    \label{fig:cat_noise_level_hyper_selection}
\end{figure*}
\section{Additional Implementation Details}
\label{appendix:additional_implementation}

This section elaborates on the hyperparameter configurations, stabilization mechanisms, and architectural variants used in our experiments.

\paragraph{Alternative Model Configurations}
To evaluate the robustness of our method, we provide comparisons between different residual prediction modes and mask rate schedules:

\begin{itemize}
    \item \textbf{Residual Prediction Mode:} This configuration dictates how the input is subtracted from the network output to formulate the denoising target.
    \begin{itemize}
        \item \textit{Adaptive (Default):} Uses the optimal LMMSE coefficient derived in Appendix \ref{appendix:alpha_opt_derivation}. The network predicts a residual where the input is scaled by a time-dependent factor $\alpha_t$: 
        \[ \mathbf{x}_{\text{out}} = F_\theta(\mathbf{x}_{\text{in}}) - \alpha_t \mathbf{x}_{\text{in}}, \quad \text{where } \alpha_t = \frac{\sigma_d t}{\sigma_d^2 + t^2}. \]
        \item \textit{Constant:} A simplified residual connection where the input is subtracted directly without scaling (effectively setting $\alpha_t = 1$). This forces the network to predict the pure residual $\mathbf{x}_{\text{out}} = F_\theta(\mathbf{x}_{\text{in}}) - \mathbf{x}_{\text{in}}$.
    \end{itemize}

    \item \textbf{Mask Rate Schedule:} This determines the mapping from the continuous noise level $t$ to the discrete mask rate $m(t) \in [0, 1]$.
    \begin{itemize}
        \item \textit{Log-uniform :} Linearly maps the logarithmic noise level $\ln(t)$ to the mask rate $m(t)$. This ensures that the discrete masking process is uniformly distributed with respect to the logarithmic time scale used by the continuous diffusion.
        \item \textit{Sigmoid-style (Default):} Adopts a sigmoid-like mapping derived from the ratio of noise to data weight, commonly used in discrete flow matching to map unbounded diffusion time $t \in [0, \infty)$ to a bounded probability:
        \[ m(t) = \frac{t}{t+\sigma_d}. \]
    \end{itemize}
\end{itemize}

\paragraph{Training and Sampling Dynamics}
We incorporate two specific mechanisms to stabilize training and refine the generation process:

\begin{itemize}
    \item \textbf{Categorical Loss Weighting:} While heuristic objectives like MaskGIT~\cite{chang2022maskgit} have shown promise, they often lack the scaling terms required for a rigorous link to maximum likelihood estimation~\cite{nie2025large}. To bridge this gap and balance the training objective, we employ a time-dependent weighting scheme inversely proportional to the mask rate:$$w_{\text{cat}}(t) = \min\left(\frac{1}{m(t)}, 10\right).$$
    By incorporating this $1/m(t)$ factor, we ensure that time steps with fewer masked tokens (low $m(t)$) are not under-represented in the total loss. The clipping mechanism further prevents numerical instability near $t=0$, effectively combining theoretical motivation with training stability.
    
    \item \textbf{Sampling Temperature:} During the generation phase, we utilize a temperature parameter $\tau$ to rescale the logits of the categorical distributions (atom types, charges, and bond types) prior to the Softmax operation: $P(z_i) = \text{Softmax}(l_i / \tau)$, where $l_i$ are the predicted logits. This parameter controls the stochasticity of the discrete transitions, balancing diversity with structural validity.
\end{itemize}

\section{Molecule Algorithm}
\label{appendix:sampling_algorithms}

This section details the sampling algorithms used for generating 3D molecules. We employ two distinct samplers: a standard diffusion stochastic sampler for models handling only continuous coordinates (Algorithm \ref{alg:sampling}), and an extended masked diffusion sampler for our primary model that jointly generates continuous coordinates and discrete atomic features (Algorithm \ref{alg:masked_sampling}).

Algorithm \ref{alg:sampling} outlines the standard stochastic sampler for a diffusion model operating on continuous variables, such as atomic coordinates $\mathcal{M}$. The process begins by sampling an initial state from a Gaussian distribution at the maximum noise level $t_0$ (Line 2). The core of the algorithm is an iterative loop that progressively denoises the sample. In each step, a small amount of noise is first added to the current sample $\mathcal{M}_i$ to create a perturbed state $\hat{\mathcal{M}}_i$ (Lines 5-6). This stochastic step (often called "churn") helps in error correction and exploration. The model $f_\theta$ then predicts the denoised version of $\hat{\mathcal{M}}_i$, from which a denoising direction $d_i$ is computed (Line 7). Finally, the sample is updated by taking a step in this direction to arrive at the less noisy state $\mathcal{M}_{i+1}$ (Line 8). This process is repeated until the noise level reaches zero, yielding the final molecular structure.

Algorithm \ref{alg:masked_sampling} extends this framework to a multi-modal scenario, jointly generating continuous coordinates ($\mathbf{x}$) and discrete atomic types ($\mathbf{z}$). The process similarly starts from pure noise for both modalities (Line 2), with atom types initialized from a uniform categorical distribution. The key distinction lies in the iterative update loop. Both coordinates and atom types are corrupted with modality-specific noise (Lines 6-7); coordinates receive Gaussian noise, while atom types are randomly masked by mixing with a uniform distribution. The model $f_\theta$ takes this combined noisy input and jointly predicts the clean coordinates and atom types (Line 8). While the coordinate update follows the same procedure as in Algorithm \ref{alg:sampling} (Lines 9-10), the discrete atom types are updated using a specialized \texttt{Discrete\_Sampler} (Line 11). This sampler is designed to handle categorical variables and implements a discrete flow matching (DFM) process, ensuring that the updates for atomic types are valid and effective. The final output is a complete 3D molecule with both its geometry and chemical composition specified.

\begin{algorithm}[h]
\caption{Diffusion Stochastic Sampler}
\label{alg:sampling}
\begin{algorithmic}[1]
    \STATE {\bfseries Input:} Sampling steps $t_0 > \cdots > t_N = 0$, model $f_\theta(\mathcal{M}, t)$, noise scale $S_{\text{noise}}$
    \STATE Sample $\mathcal{M} \sim \mathcal{N}(0, t_0^2 I)$
    \FOR{$i = 0$ to $N-1$}
        \STATE Sample $\epsilon_i \sim \mathcal{N}(0, I)$ 
        \STATE $\hat{t}_i \gets t_i + \gamma_i t_i$ 
              
        \STATE $\hat{\mathcal{M}}_i \gets \mathcal{M}_i + \sqrt{\hat{t}_i^2 - t_i^2} \cdot \epsilon_i$
        
        \COMMENT{$\triangleright$ Increased noise level from $t_i$ to $\hat{t}_i$}

        \STATE $d_i \gets (\hat{\mathcal{M}}_i - f_\theta(\hat{\mathcal{M}}_i, \hat{t}_i)) / \hat{t}_i$
        \STATE $\mathcal{M}_{i+1} \gets \hat{\mathcal{M}}_i + (t_{i+1} - \hat{t}_i) \cdot d_i$
    \ENDFOR
    \STATE {\bfseries Output:} $x_N$
\end{algorithmic}
\end{algorithm}

\begin{algorithm}[h]
\caption{Masked Diffusion Stochastic Sampler for Joint Coordinate and Atom-Type Generation}
\label{alg:masked_sampling}
\begin{algorithmic}[1]
    \STATE {\bfseries Input:} Sampling steps $t_0 > \cdots > t_N = 0$, model $f_\theta(\{\textbf{x},\textbf{z}\}, t)$, mask\_rate function $m(t)$, number of category S
    \STATE Sample $\mathbf{x}_0 \sim \mathcal{N}(0, t_0^2 I)$, $\mathbf{z}_0 \sim \mathrm{Cat}\left(\frac{1}{S}, \dots, \frac{1}{S}\right)$
    \FOR{$i = 0$ to $N-1$}
        \STATE Sample $\boldsymbol{\epsilon}_i \sim \mathcal{N}(0, I)$ 
        \STATE $\hat{t}_i \gets t_i + \gamma_i t_i$ 
        \STATE  $\hat{\mathbf{x}}_i \gets \mathbf{x}_i + \sqrt{\hat{t}_i^2 - t_i^2} \cdot \boldsymbol{\epsilon}_i$ 
        \STATE$\hat{\mathbf{z}}_i \sim Cat[(1-r_i) \cdot \delta_{\mathbf{z}_i} + r_i \cdot \mathrm{Unif}(1, S)]$
        \STATE $\mathbf{x}_{\mathrm{pred}}, \mathbf{z}_{\mathrm{pred}} \gets f_\theta(\{\hat{\mathbf{x}}_i,\hat{\mathbf{z}}_i\}, \hat{t}_i)$
        \STATE $\mathbf{d}_i \gets (\hat{\mathbf{x}}_i -\mathbf{x}_{\mathrm{pred}}) / \hat{t}_i$
        
        \STATE $\mathbf{x}_{i+1} \gets \hat{\mathbf{x}}_i + (t_{i+1} - \hat{t}_i) \cdot \mathbf{d}_i$

        \STATE $\mathbf{z}_{i+1} \gets \text{Discrete\_Sampler}(\hat{\mathbf{z}}_i, \mathbf{z}_{\mathrm{pred}}, m(\hat{t}_i) - m(t_{i+1}))$ \hfill 

    \ENDFOR
    \STATE {\bfseries Output:} $(\mathbf{x}_N,\mathbf{z}_N)$
\end{algorithmic}
\end{algorithm}

\section{Experiments of various sampling steps}
As shown in Table~\ref{tab:geom_drugs_xtb_full}, our VEDA-S model consistently achieves state-of-the-art performance across most metrics, even at a low number of sampling steps (e.g., 100), outperforming prior methods like SemlaFlow and FlowMol2 in terms of molecule stability and RMSD.

\begin{table*}[ht]
\centering
\small
    
    \begin{tabular}{l|c|cc|ccc|ccll}
        \hline
        Model & NFE& MS$\uparrow$ & \makecell{V\&C}$\uparrow$ & \makecell{Bond\\Length\\($\times10^{-2}$)}$\downarrow$ & \makecell{Bond\\Angles}$\downarrow$ & Torsions$\downarrow$ & \makecell{Median\\$\Delta E_{relax}$}$\downarrow$ & \makecell{Mean\\$\Delta E_{relax}$}$\downarrow$  &\makecell{Median\\RMSD}$\downarrow$ &\makecell{Mean\\RMSD}$\downarrow$  \\
        \hline

 EQGAT & 250& 0.318& 0.605& 2.63& 3.01& 17.63& 34.95& 76.49& 0.974&1.044\\
        EQGAT &500 & 0.899 & 0.834 & 1.00& 1.15& 8.58& 6.40& 11.1  &0.915&0.975\\
        JODO &500 & 0.963 & 0.879 & 0.77& 0.83& 6.01& 4.74& 7.04  &-&-\\
        Megalodon-quick &500 & 0.944 & 0.900& 0.66& 0.71& 5.58& 3.19& 5.76  &-&-\\
        \hline
        FlowMol2 &50 & 0.750& 0.694& 1.87& 1.99& 16.22& 23.9&31.34&0.947& 1.024\\
        FlowMol2 &100 & 0.830& 0.768& 1.54& 1.78& 15.45& 21.0& 27.1&0.992&1.038\\
        FlowMol2 &150 & 0.883& 0.796& 1.37& 1.65& 14.90& 20.1&25.8&0.950& 1.011\\
        FlowMol2 &200 & 0.902& 0.803& 1.26& 1.61& 14.70& 16.9&23.5&0.981& 0.999\\
        FlowMol2 &250 & 0.892& 0.807& 1.30& 1.60& 14.59& 17.8&23.9&0.954& 1.007\\
        
        SemlaFlow &50 & 0.965& 0.899& 2.97& 2.02& 6.52& 33.2& 93.6&0.310&0.433\\
        SemlaFlow &100 & 0.969& 0.920& 3.10& 2.06& 6.05& 32.3& 91.0  &0.273&0.358\\
        SemlaFlow &150 & 0.971&  0.943& 2.88& 1.92& 5.11& 31.5& 68.7&0.230&0.349\\
        SemlaFlow &200 & 0.974& 0.942& 2.86& 1.90& 4.93& 31.4& 90.4&0.227&0.355\\
        SemlaFlow &250 & 0.978& 0.933& 2.84& 1.87& 4.86& 30.4& 95.0&0.216&0.338\\
        Megalodon-flow &100 & 0.987& 0.948& 2.30& 1.62& 5.58& 20.9& 46.9  &-&-\\
        \hline
VEDA-S  &50 & \makecell{0.968\\\scriptsize$\pm$0.001} & \makecell{0.966\\\scriptsize$\pm$0.010} & \makecell{0.86\\\scriptsize$\pm$0.02} & \makecell{0.60\\\scriptsize$\pm$0.04} & \makecell{5.10\\\scriptsize$\pm$0.39} & \makecell{2.99\\\scriptsize$\pm$0.34} & \makecell{8.38\\\scriptsize$\pm$2.05} & \makecell{0.236\\\scriptsize$\pm$0.055} & \makecell{0.355\\\scriptsize$\pm$0.042} \\
VEDA-S  &75 & 0.992 & 0.981 & 0.73 & 0.46 & 3.38 & 2.06 & 6.66 & 0.138 & 0.262 \\
VEDA-S &100 & \makecell{0.995\\\scriptsize$\pm$0.001} & \makecell{0.988\\\scriptsize$\pm$0.003} & \makecell{0.69\\\scriptsize$\pm$0.02} & \makecell{0.41\\\scriptsize$\pm$0.02} & \makecell{2.71\\\scriptsize$\pm$0.25} & \makecell{1.72\\\scriptsize$\pm$0.17} & \makecell{2.93\\\scriptsize$\pm$0.13} & \makecell{0.096\\\scriptsize$\pm$0.02} & \makecell{0.197\\\scriptsize$\pm$0.03} \\
VEDA-S &125 & 0.999& 0.986& 0.69& 0.38& 2.24& 1.61&2.55&0.084&0.161\\
VEDA-S &150 & 0.995& 0.991& 0.66& 0.36& 1.94& 1.43&2.16&0.073&0.131\\
VEDA-S &200 & 0.999& 0.993& 0.65& 0.34& 1.68& 1.35&1.99&0.064&0.114\\
VEDA-S &250 & 0.998& 0.987& 0.67& 0.34& 1.59& 1.34&2.06&0.061&0.107\\
         \hline
    \end{tabular}
\caption{Performance comparison on the GEOM-DRUGS dataset following the benchmark of \citet{nikitin2025geom}. Results for FlowMol2, EQGAT, SemlaFlow, and VEDA-S are reported under different sampling steps. VEDA-S is evaluated using 5000 generated samples at 50 and 100 steps, with standard deviations computed over 5 subsets. All other methods are evaluated on 1,000 generated samples.}
\label{tab:geom_drugs_xtb_full}
\end{table*}

\section{Derivation of Transition Rates Matrix for the Masking Process}
\label{appendix:mask_transition_rate}
Following the derivation of Discrete Flow Matching~\cite{campbell2024generative}, we will derive the form $R^*$, representing the transition rate from the current state $x_t$ to each state $j$, with respect to $t$ and $x_0$
Let the number of categories be $S$, and the mask rate function satisfy $m(t)\in[0,1]$ with $m'(t)>0$. Denote by
\[
\delta(i, x_0) =
\begin{cases}
1, & i = x_0,\\
0, & i \neq x_0,
\end{cases}
\]
the Kronecker delta. We interpolate between the point mass at $x_0$ and the uniform distribution:
\begin{equation}
p(x_t)
= \mathrm{Cat}\bigl((1 - m(t))\,\delta(x_t, x_0) + \tfrac{m(t)}{S}\bigr).
\end{equation}

\paragraph{Predictive Term} Define the instantaneous transition rate from state $x_t$ to state $j$ as
\begin{equation}
R^*(x_t, j)
= \frac{\mathrm{ReLU}\bigl(\partial p(j) - \partial p(x_t)\bigr)}{S\,p(x_t)}.
\end{equation}

Since
\begin{align}
\partial p(x_t) &= \partial\bigl((1 - m(t))\,\delta(x_t, x_0) + \tfrac{m(t)}{S}\bigr)\\
&= -m'(t)\,\delta(x_t, x_0) + \frac{m'(t)}{S}
\end{align}

we obtain
\begin{equation}
R^*(x_t,j)
= -\frac{\mathrm{ReLU}\bigl(\delta(j, x_0) -\delta(x_t, x_0) \bigr )\,m'(t)}
         {S\bigl((1 - m(t))\,\delta(x_t, x_0) + \tfrac{m(t)}{S}\bigr)}.
\end{equation}

Noting that $R^*(x_t,j)\neq0$ if and only if $x_t = j \neq x_0$, we simplify:
\begin{align}
R^*(x_t,j)
&= \frac{m'(t)}{m(t)}\,\delta(j, x_0)\,\bigl(1 - \delta(x_t, x_0)\bigr)\\
\end{align}

\paragraph{Detailed Balanced Term} The concept of detailed balance has been utilized in previous continuous-time Markov chain (CTMC) generative models, such as in the work by \citet{campbell2024generative}, to make adjustments for inference. We add a detailed-balance correction satisfying
\begin{equation}
p(i)\,R^{\mathrm{DB}}(i,j)
= p(j)\,R^{\mathrm{DB}}(j,i),
\end{equation}

We need to make some assumptions about the form of $R^{\mathrm{DB}}$. When considering masking noise, a process that is in detailed balance will have a certain rate for transitions from a masked state to $z_0$, as well as a certain rate for transitions from $z_0$ back to the masked state in order to cancel out this effect. The rate for such transitions would take the following form. So We have parameterized $R^{\mathrm{DB}}$ with two time-dependent pre-defined constants $a_t$ and $b_t$. Therefore, when $ i $ and $ j $ are both not equal to $z_0 $, this term is zero, essentially representing the diffusion flow between $ z_0 $ and other states and its reverse action.

\begin{equation}
R^{\mathrm{DB}}(i,j) = a_t\,\delta(i, z_0) + b_t\,\delta(j, z_0).
\end{equation}
 
\begin{equation}
p(i)=(1-m(t))\,\delta(i, z_0)+\frac{m(t)}{S}
\end{equation}

This yields \begin{align}
&\bigl((1-m)\,\delta(i, z_0)+\tfrac{m}{S}\bigr)
\bigl(a_t\,\delta(i, z_0)+b_t\,\delta(j, x_0)\bigr)\\
=&\bigl((1-m)\,\delta(j, z_0)+\tfrac{m}{S}\bigr)
\bigl(a_t\,\delta(j, z_0)+b_t\,\delta(i, z_0)\bigr).
\end{align}

For the case $x_0 = i$, $\delta(i, z_0)=1$ and $\delta(j, z_0)=0$, we have:
\begin{align}
&\bigl(1 - m(t) + \tfrac{m(t)}{S}\bigr)\,a_t
= \tfrac{m(t)}{S}\,b_t\\
&b_t = \frac{S - S\,m(t) + m(t)}{m(t)}\,a_t.
\end{align}

Introducing a noise strength parameter $\eta$, we set
\begin{equation}
R^{\mathrm{DB}}(i,j)
= \eta\,\delta(i, z_0)
+ \eta\,\frac{S - S\,m(t) + m(t)}{m(t)}\,\delta(j, x_0).
\end{equation}

\paragraph{Full Transition Rate} The full transition rate $R^{\theta}$ consists of two components: the base transition $R^*$, which arises from the interpolation between a point mass at the original category and a uniform distribution, and the correction term $R^{\mathrm{DB}}$, which enforces detailed balance to ensure proper Markovian dynamics. The total rate is given by:
\begin{align}
R^{\theta}(z_t,j) &= 
\mathbb{E}_{p^{\theta}_{0|t}(z_0 \mid z_t)}
\left[
R^*(z_t,j) + R^{\mathrm{DB}}(z_t,j)
\right]\\
&= 
\mathbb{E}_{p^{\theta}_{0|t}(z_0 \mid z_t)}
\left[
\frac{m'(t)}{m(t)}\,\delta(j, z_0)\bigl(1- \delta(z_t, z_0) \bigr) \right.\notag\\
&\qquad\left. +\eta\,\delta(z_t, z_0)
+ \eta\,\frac{S - S\,m(t) + m(t)}{m(t)}\,\delta(j, z_0)
\right]\\
&=\frac{\eta S(1- m(t)) + \eta m(t) +m'(t)}{m(t)}p^{\theta}_{0|t}(z_0 = j \mid z_t) \notag\\
&\qquad + \eta p^{\theta}_{0|t}(z_0 = z_t \mid z_t)
\end{align}

In this formulation, the term $p^{\theta}_{0|t}(z_0 = j \mid z_t)$ represents the predicted probability that the clean data corresponds to category $j$, and thus determines the transition rate from the current state $z_t$ to $j$. The second term, $p^{\theta}_{0|t}(z_0 = z_t \mid z_t)$, represents the probability that the current noisy state $z_t$ already matches the clean data, and therefore governs the rate of remaining in the current state.

\paragraph{Transition Rate under Our Mask Rate Schedule}
With our specific mask rate schedule 
\[m(t) = \frac{\log(t/T_{\min})}{\log(T_{\max}/T_{\min})},\]
we can derive the explicit form of the transition rate. Taking the derivative with respect to time, we obtain 
\[m'(t) = \frac{1}{t\log(T_{\max}/T_{\min})}.\]
\begin{align}
R &= \left[\eta S \frac{\log(T_{\max}/T_{\min})}{\log(t/T_{\min})} + \eta(1-S) + \frac{1}{t\log(t/T_{\min})}\right]\\
&p^{\theta}_{0|t}(x_0 = j \mid x_t) + \eta p^{\theta}_{0|t}(x_0 = x_t \mid x_t)
\end{align}

In our formulation, we replace the original time variable $t$ in the base paper with a monotonic schedule function of mask rate $m(t)$. To match the effective categorical noise strength used in the original Discrete Flow Matching formulation (where $t$ directly corresponds to the interpolation coefficient), one could compensate by multiplying the noise parameter $\eta$ by $m'(t)$. However, since the detailed balance condition holds for any constant scaling of $\eta$, this modification is not strictly required. In practice, we treat the inclusion or exclusion of the extra $m'(t)$ factor as a tunable hyperparameter, and adopt the version that includes it in our final implementation.

\section{Derivation of Transition Rates Matrix for the Masking Process}
\label{appendix:gat_dfm_mask_transition_rate}

We also include the transition rate matrix formulation from Discrete Flow Matching (DFM) \cite{gat2024discrete}, which provides an alternative approach to modeling discrete diffusion dynamics.
This approach offers computational efficiency through streamlined probability updates, making it particularly suitable for large-scale discrete token generation. 
The DFM approach constructs a velocity field using forward and backward components:
\begin{equation}
\bar{u}_t = \alpha_t \hat{u}_t - \beta_t \tilde{u}_t
\end{equation}
where $\hat{u}_t = \frac{m'(t)}{m(t)}(p_\theta - p_{\text{curr}})$ represents the forward velocity toward model predictions, $\tilde{u}_t = \frac{m'(t)}{1-m(t)}(p_{\text{curr}} - p_{\text{noise}})$ represents the backward velocity, and $\alpha_t$ and $\beta_t$ are time-dependent coefficients satisfying $\alpha_t + \beta_t = 1$. More specifically, $\alpha_t \propto \eta $, where $\eta$ is the categorical noise level. The discrete transition probabilities are then updated as:
\begin{equation}
p_{\text{next}} = p_{\text{curr}} + \Delta m(t) \cdot \bar{u}_t
\end{equation}
\paragraph{DFM Formulation}
Rather than directly computing transition rates, DFM constructs a velocity field using a mask rate schedule $m(t)$.  The approach decomposes the dynamics into two velocity components:
\begin{itemize}
    \item \textbf{Forward velocity}: $\hat{u}_t = \frac{m'(t)}{1-m(t)}(p_\theta - p_{\text{curr}})$, which drives transitions from the current state toward the model's predicted distribution
    \item \textbf{Backward velocity}: $\tilde{u}_t = \frac{m'(t)}{m(t)}(p_{\text{curr}} - p_{\text{noise}})$, which accounts for the reverse flow from the current state toward the noise distribution
\end{itemize}

\paragraph{Velocity Field Combination}

The final velocity field is constructed as a weighted combination:
$$\bar{u}_t = \alpha_t \hat{u}_t - \beta_t \tilde{u}_t$$

where the weighting parameters satisfy $\alpha_t + \beta_t = 1$ and are typically parameterized as:
$\alpha_t = \eta (1-m(t))^a m(t)^b$

with $\eta$, $a$, and $b$ being hyperparameter that control the relative importance of forward and backward flows throughout the process.

\paragraph{Transition Rate Matrix}

The discrete transition probabilities are updated using:
$p_{\text{next}} = p_{\text{curr}} + \Delta m(t) \cdot \bar{u}_t$

This formulation provides a direct way to compute transition rates by combining model predictions with the current state distribution, weighted by the time-dependent schedule $m(t)$ and its derivative $m'(t)$.

\paragraph{Comparison with Standard Diffusion}
Unlike our direct derivation of $R^*(x_t, j)$ from the interpolation dynamics, DFM's approach allows for more flexible control over the masking dynamics through the choice of $m(t)$ and the hyperparameters governing $\alpha_t$. Notably, DFM's formulation is conceptually simpler than the \cite{campbell2024generative} framework, as it avoids special treatment of diagonal elements (i.e., no-transition cases) in the transition matrix. While DFM introduces some empirically-motivated hyperparameters, this simplicity provides better interpretability for downstream applications and facilitates debugging during implementation. The explicit separation into forward and backward components also provides clear intuition about the direction of information flow at each time step.

Although the backward velocity field in DFM serves to balance model predictions and can provide effects to avoid overly greedy sampling, our experiments reveal that it cannot fully substitute for explicit remasking and noise injection strategies, likely due to the multi-modal nature of molecular structures. Both methods enable iterative refinement of masked regions and improve the consistency and quality of generated discrete sequences.

\section{Derivation of Optimal $\alpha_t$}
\label{appendix:alpha_opt_derivation}
In the main text, we propose to decouple the network's learning task by isolating and subtracting the undesired linear component that arises from the network's inductive bias toward identity-like functions. We formalize this by finding a coefficient $\alpha_t$ that provides the best linear estimate of the target noise $\sigma_d \boldsymbol{\epsilon}$ given the noisy input $\mathbf{x}_t$.

This coefficient is the solution to the following Linear Minimum Mean Squared Error (LMMSE) problem:
\begin{equation}
    \alpha_t = \arg \min_\alpha\ \mathbb{E}[\| \sigma_d \boldsymbol{\epsilon} - \alpha \mathbf{x}_t \|^2] \label{eq:lmmse}
\end{equation}

The LMMSE framework is chosen because it provides a principled and optimal solution for this specific goal. Its use rests on two key premises. First, we deliberately constrain the estimator to be \textbf{linear} (i.e., of the form $\alpha \mathbf{x}_t$), which directly corresponds to our objective of quantifying and removing the network's linear bias. Second, we assume the \textbf{second-order statistics} (i.e., variances and covariances) of the random variables $\mathbf{x}_t$ and $\boldsymbol{\epsilon}$ are known. This assumption holds true in diffusion models, where these statistics are well-defined functions of the noise level $t$ and data properties. The LMMSE formulation in Eq.~\eqref{eq:lmmse} is the standard method for finding this optimal linear estimator.

To find the minimum, we differentiate the objective function $J(\alpha) = \mathbb{E}[\| \sigma_d \boldsymbol{\epsilon} - \alpha \mathbf{x}_t \|^2]$ with respect to $\alpha$ and set the result to zero:
\begin{align}
    \frac{\partial J(\alpha)}{\partial \alpha} &= \frac{\partial}{\partial \alpha} \mathbb{E}\left[ (\sigma_d \boldsymbol{\epsilon})^T(\sigma_d \boldsymbol{\epsilon}) - 2\alpha (\sigma_d \boldsymbol{\epsilon})^T \mathbf{x}_t + \alpha^2 \mathbf{x}_t^T \mathbf{x}_t \right] \\
    &= \mathbb{E}\left[ -2 (\sigma_d \boldsymbol{\epsilon})^T \mathbf{x}_t + 2\alpha \mathbf{x}_t^T \mathbf{x}_t \right] = 0
\end{align}
Solving for $\alpha$ yields:
\begin{equation}
    \alpha_t = \frac{\mathbb{E}[(\sigma_d \boldsymbol{\epsilon})^T \mathbf{x}_t]}{\mathbb{E}[\mathbf{x}_t^T \mathbf{x}_t]} = \frac{\text{Cov}(\sigma_d \boldsymbol{\epsilon}, \mathbf{x}_t)}{\text{Var}(\mathbf{x}_t)}
\end{equation}
where the second equality holds for zero-mean variables.

Following the standard diffusion model setup, the noisy sample is defined as $\mathbf{x}_t = \mathbf{x}_0 + t \boldsymbol{\epsilon}$, where the clean data $\mathbf{x}_0$ and the standard Gaussian noise $\boldsymbol{\epsilon} \sim \mathcal{N}(0, \mathbf{I})$ are independent. We adopt the setup from \citet{karras2022elucidating}, where $\text{Var}(\mathbf{x}_0) = \sigma_d^2$ and $\text{Var}(\boldsymbol{\epsilon}) = 1$.

\paragraph{Covariance (Numerator):}
\begin{align*}
    \text{Cov}(\sigma_d \boldsymbol{\epsilon}, \mathbf{x}_t) &= \text{Cov}(\sigma_d \boldsymbol{\epsilon}, \mathbf{x}_0 + t \boldsymbol{\epsilon}) \\
    &= \sigma_d \text{Cov}(\boldsymbol{\epsilon}, \mathbf{x}_0) + \sigma_d t \text{Var}(\boldsymbol{\epsilon}) \\
    &= 0 + \sigma_d t \cdot 1 = \sigma_d t
\end{align*}

\paragraph{Variance (Denominator):}
\begin{align*}
    \text{Var}(\mathbf{x}_t) &= \text{Var}(\mathbf{x}_0 + t \boldsymbol{\epsilon}) \\
    &= \text{Var}(\mathbf{x}_0) + t^2 \text{Var}(\boldsymbol{\epsilon}) \\
    &= \sigma_d^2 + t^2
\end{align*}

Substituting these results gives the final expression:
\begin{equation}
    \alpha_t = \frac{\sigma_d t}{\sigma_d^2 + t^2}
\end{equation}
This choice of $\alpha_t$ removes the optimal linear estimate of the noise from the network's output, thereby allowing the network to focus on learning the non-linear residual component.

\section{Derivation for the Smoothed Potential Energy Equivalence}
\label{appendix:smoothing_derivation}

Below, we provide a formal derivation for the claim made in the main text that adding Gaussian noise to coordinates is equivalent in expectation to sampling from a smoothed potential energy surface. We demonstrate this for the 3D case, which is relevant to molecular structures.

Let the molecular coordinates be $\mathbf{x} \in \mathbb{R}^3$ and the potential energy surface (PES) be the function $E(\mathbf{x})$.

First, let's define the smoothed PES, $\tilde{E}(\mathbf{x})$, as the convolution of the original surface $E(\mathbf{x})$ with a 3D Gaussian kernel $G(\mathbf{z}; \sigma^2)$:
\begin{equation}
    \tilde{E}(\mathbf{x}) = \int_{\mathbb{R}^3} E(\mathbf{y}) G(\mathbf{x} - \mathbf{y}; \sigma^2) d\mathbf{y}
    \label{eq:appendix_smoothed_def}
\end{equation}
where $G(\mathbf{z}; \sigma^2)$ is the standard zero-mean Gaussian probability density function with variance $\sigma^2$ along each axis.

Our procedure involves adding Gaussian noise $\bm{\epsilon} \sim \mathcal{N}(\mathbf{0}, \sigma^2 I)$ to the coordinates, resulting in a perturbed state $\mathbf{x}' = \mathbf{x} + \bm{\epsilon}$. Our goal is to show that the expected energy of this new state is exactly the value of the smoothed potential at the original location $\mathbf{x}$, which can be expressed by:
\begin{equation}
\mathbb{E}_{\bm{\epsilon}}[E(\mathbf{x} + \bm{\epsilon})] = \tilde{E}(\mathbf{x})    
\end{equation}

We start from the definition of the expectation:
\begin{align*}
    \mathbb{E}_{\bm{\epsilon}}[E(\mathbf{x} + \bm{\epsilon})] &= \int_{\mathbb{R}^3} E(\mathbf{x} + \bm{\epsilon}) p(\bm{\epsilon}) d\bm{\epsilon} \\
    &= \int_{\mathbb{R}^3} E(\mathbf{x} + \bm{\epsilon}) G(\bm{\epsilon}; \sigma^2) d\bm{\epsilon}
\end{align*}
Here, $p(\bm{\epsilon})$ is the probability density function of the noise, which is the Gaussian kernel $G$.

Next, we perform a simple change of variables by setting $\mathbf{y} = \mathbf{x} + \bm{\epsilon}$. This implies $\bm{\epsilon} = \mathbf{y} - \mathbf{x}$. Substituting this back into the integral gives:
\begin{equation}
\mathbb{E}_{\bm{\epsilon}}[E(\mathbf{x} + \bm{\epsilon})] = \int_{\mathbb{R}^3} E(\mathbf{y}) G(\mathbf{y} - \mathbf{x}; \sigma^2) d\mathbf{y}
\end{equation}
Recognizing that the Gaussian kernel is symmetric, i.e., $G(\mathbf{y} - \mathbf{x}) = G(\mathbf{x} - \mathbf{y})$, the right-hand side of the equation becomes identical to the definition of the smoothed potential $\tilde{E}(\mathbf{x})$ in Equation~\ref{eq:appendix_smoothed_def}.

Thus, we have shown that:
\begin{equation}
    \mathbb{E}_{\bm{\epsilon}}[E(\mathbf{x} + \bm{\epsilon})] = \tilde{E}(\mathbf{x}) \quad 
\end{equation}
This confirms that, on average, the energy of the noise-perturbed samples corresponds to the energy of a smoothed landscape, which facilitates exploration and helps the sampler avoid minor local minima.

\else
\fi

\end{document}